\documentclass[11pt]{JHEP3}
\usepackage{amsmath}
\usepackage{amssymb}
\usepackage{bbold}
\DeclareMathOperator{\tr}{tr}
\DeclareMathOperator{\Real}{Re}
\DeclareMathOperator{\Imag}{Im}
  
\title{Diagonal D-branes in product spaces and their Penrose limits} 

\author{Gor Sarkissian and Marija Zamaklar\\
The Abdus Salam Centre ICTP,\\
Strada Costiera 11,\\
34014, Trieste, ITALY
} 
\preprint{hep-th/0308174}
\date{August 26th, 2003} 

\abstract{We study classes of D-branes embedded in various
$\text{AdS}^{m}\times S^n \times S^p \times T^{q}$ backgrounds, which
nontrivially mix the target-space submanifolds. Mixing is achieved
either via diagonal geometric embedding or through a mixed worldvolume
flux which has one index in the sphere and one index in the AdS part.
Branes of the former type wrap calibrated cycles in the target space,
while those of the latter type wrap \emph{non-supersymmetric} target
space cycles which are stabilised only after the mixed worldvolume
flux is turned on.
In the second part of the paper we study two qualitatively different
Penrose limits of these diagonal branes. In the first case we look at
geodesics which \emph{do not} belong to the worldvolume of brane. In
order to get a nontrivial result, one needs to bring the brane closer
and closer to the geodesic while taking the limit. The result is a
D-brane with a worldvolume relativistic pulse.
In the second case the Penrose geodesic belongs to the worldvolume and
the resulting brane is of the ``oblique'' type: it is diagonally
embedded between different~$SO$ groups of the target space pp-wave.}

\begin{document}
\section{Introduction and summary}

The study of D-branes in curved space hardly needs any further
justification: from the pure CFT point-of-view they represent an
interesting arena for the study of various CFT phenomena, they are
crucial for the understanding of non-perturbative effects in the
context of the AdS/CFT correspondence, and they are essential
ingredients for the realisation of AdS/defect (conformal) field theory
dualities.

In this paper we continue the study of D-branes in different curved
manifolds. In the first part of the paper we address the issue of
D-branes in manifolds which are products of several manifolds, while
the branes themselves are \emph{not} products of branes existing on
these submanifolds. Instead, the branes nontrivially mix the target
space submanifolds. Mixing is achieved via: 1)~a \emph{diagonal
geometric embedding} or 2)~through a \emph{mixed worldvolume flux}
which has one index in the sphere and one index in the AdS part,
i.e.~with a field strength of the schematic form~$F_{\text{sphere},
\text{AdS}}$.

One way to construct the first class of branes is to start with
supersymmetric flat space configurations of branes intersecting under
angles.  Replacing some of the branes by their supergravity solutions
and subsequently focusing on their corresponding near-horizon
geometries, while keeping the remaining branes as probes, one can
derive the effective geometries for the probes. They nontrivially mix
all submanifolds of the near horizon space. Because of their flat
space origin we call them \emph{diagonal} branes. Some of the diagonal
branes have previously appeared in the literature. The diagonal~$D3$
and~$D5$ in $\text{AdS}_5 \times S^5$ have been constructed
\cite{Bilal:1998ck,Gutowski:1999iu}, while various diagonal $M2$ and
$M5$ branes in $\text{AdS}_4\times S^7$ have been considered
\cite{Gutowski:1999tu} in the context of generalised calibrations. In
our study we however focus on different kinds of backgrounds.

In our study of D-branes in AdS spaces we mainly restrict to target
spaces that are products of \emph{group} manifolds: $\text{AdS}_3
\times S^3 \times T^4$ and $\text{AdS}_3 \times S^3 \times S^3 \times
R$. In the first manifold, we consider a brane which mixes \emph{all
submanifolds} and interpolates between a brane with embedding
$\text{AdS}_2 \times S^2\times{\rm point}$ and a brane with embedding
$\text{AdS}_2 \times {\rm point}\times T^2$.  In the second target
space we consider a brane which interpolates between a brane with
embedding $\text{AdS}_2 \times S^2 \times \text{point} \times R$ and a
brane with embedding $\text{AdS}_2 \times \text{point} \times S^2
\times R$. We derive the CFT symmmetries preserved by these branes,
but due to the low number of preserved symmetries the construction of
boundary states appears to be a hard problem.  However, while a direct
CFT analysis of these geometric diagonal branes does not look too
tractable, an alternative and apparently more manageable approach is
the one which uses the dCFT/AdS correspondence. Diagonal branes should
correspond to RG-flows between \emph{different} boundary conformal
field theories. Especially interesting configurations are those of
diagonal $D3$ and $D5$ branes in $\text{AdS}_5\times S^5$, which
deserve a separate study~\cite{sarkissian}.

The first class of branes (which mixes submanifolds through a diagonal
geometric embedding) can be described using
calibration techniques.  
There are frequent statements in the literature that the flat space
calibration results~\cite{Gibbons:1998hm} (for example that
supersymmetric branes wrap holomorphic cycles in the target space) can
be extended in a straightforward manner to curved spaces. However, we
point out that in cases when the full target space is not a complex
manifold, the these results can be applied only to particular complex
submanifolds for which we know in advance that it is consistent to
truncate the DBI action to. A simple example is provided by an
$\text{AdS}_3 \times S^3 \times T^4$ space in which we explicitly show
that the calibration of the $D3$ brane can only be applied to
submanifolds which contain a \emph{maximal} $S^2$ in $S^3$.

The construction of the second type of diagonal branes is motivated by
the \emph{flat} space T-duality between branes under angles and branes
with non-vanishing worldvolume fluxes. In curved spaces duality
generically changes the background making these flat space relations
less manifest.  We show how in some cases these flat space dualities
are inherited by the curved geometries.  Starting with the flat space
configuration of branes intersecting under an angle, we perform
T-duality in such a way that the branes which will be replaced with
the background \emph{do not} carry any worldvolume flux, while a brane
which will become a probe carries flux. Then we take a near horizon
limit of this configuration. The near horizon brane wraps a
\emph{non-supersymmetric} target space cycle (a
maximal~$\text{AdS}_4\times S^4$ in $\text{AdS}_5 \times S^5$) which
is stabilised (becomes supersymmetric) only after the mixed
worldvolume flux is turned on. This flux is different in nature from
the one considered in \cite{Bachas:2000fr,Karch:2000ct}: while the
fluxes which are considered there modify the geometry of the brane,
the flux which we consider \emph{does not}. Maximal $\text{AdS}_4
\times S^4$ is a solution to the DBI action with and without
worldvolume flux and turning on the flux does not change the extrinsic
curvature of the brane.  This distinction is a consequence of the fact
that the fluxes in \cite{Bachas:2000fr, Karch:2000ct} originate from a
brane ending on (and pulling) the other brane, while in our case it is
a consequence of T-duality.

In the second part of this paper we consider various Penrose limits of
the diagonal branes constructed in the first part.  We use two types
of geodesics. In the \emph{first case} we look at a geodesic which
\emph{does not} belong to the worldvolume of the brane. In order to
get a nontrivial result, one needs to bring the brane closer and
closer to the geodesic while taking the limit. The resulting D-brane
is a brane with a relativistic pulse propagating on its
worldvolume. These branes have recently been analysed in \emph{flat}
space in \cite{Bachas:2002jg}.  In the \emph{second case} the Penrose
geodesic belongs to the worldvolume of the diagonal brane and the
resulting brane is of the ``oblique'' type: it is diagonally embedded
between different $SO$ groups of the target space pp-wave. These kinds
of branes have been  recently discovered in~\cite{Hikida:2002qk}, and
further analysed in~\cite{Gaberdiel:2003sb}. Here
we point out their AdS origin.

At the end of the paper we also analyse the Penrose limit of branes
with diagonal flux. Generically, in order to obtain a finite flux in
the Penrose limit for these branes, one needs to simultaneously scale
the geodesic to be closer and closer to the brane surface, and require
it to become \emph{orthogonal} to the worldvolume flux. The Penrose
limit of a brane with diagonal flux leads to a D-brane with null flux.

\section{Diagonal branes from branes under the angle in flat space}

The first class of D-branes we want to study are D-branes which mix
various target space submanifolds in a \emph{geometric} way.  As
explained in the introduction these can be derived from the
(supersymmetric) flat space configurations of branes intersecting
under angles.  More precisely we consider the following supersymmetric
flat space configurations:
\begin{equation}
\label{conf1}
\begin{array}{lcccccccccc}
F1   & 0 & 1 & - & - & - & - & - & - & - & -  \\ 
NS5   & 0 & 1 & 2 & 3 & 4& 5 & - & - & - & -  \\ 
D3'  & 0 & - & - & - & - & - & 6 & 7 & 8 & -  \\
D3'' & 0 & - & 2 & 3 & - & - & 6 & - & - & -  \\ 
\end{array}
\end{equation}
and
\begin{equation}
\label{conf2}
\begin{array}{lcccccccccc}
F1   & 0 & 1 & - & - & - & - & - & - & - & -    \\
NS5  & 0 & 1 & 2 & 3 & 4 & 5 & - & - & - & -  \\ 
NS5  & 0 & 1 & - & - & - & - & 6 & 7 & 8 & 9  \\ 
D4'  & 0 & - & 2 & 3 & 4 & - & 6 & - & - & -  \\ 
D4'' & 0 &   & - & - & 4 & - & 6 & 7 & 8 & -  \\ 
\end{array}
\end{equation}
The second configuration can be derived from the eleven-dimensional
configuration of a ``non-standard'' brane intersection by dimensional
reduction in direction~$10$~\cite{Gauntlett:1997cv}:
\begin{equation}
\begin{array}{lccccccccccc}
M2   & 0 & 1 & - & - & - & - & - & - & - & - & 10   \\
M5  & 0 & 1 & 2 & 3 & 4 & 5 & - & - & - & - & -  \\ 
M5  & 0 & 1 & - & - & - & - & 6 & 7 & 8 & 9 & - \\ 
M5'  & 0 & - & 2 & 3 & 4 & - & 6 & - & - & - & 10  \\ 
M5'' & 0 &   & - & - & 4 & - & 6 & 7 & 8 & - & 10  \\ 
\end{array}
\end{equation}
The unprimed branes are subsequently replaced with their corresponding
supergravity solutions and finally with the near horizon geometries,
while the primed branes are treated as probes. A generic probe brane is
taken to be an arbitrary brane which interpolates between the two
reference (primed) branes marked in the table.  We will see that
knowing the flat space embedding of the probe branes suggests the
ansatz which solves the effective equations of motion in the near horizon
limit.

Note also that due to the way we construct branes in the near horizon
geometries (namely starting from physically acceptable brane
configurations in flat space), we are guaranteed that the final brane
configurations are ``good'' branes, i.e.~that they give rise to consistent
CFTs.

\subsection{Diagonal $D3$ brane in the $\text{AdS}_3 \times S^3 \times T^4$ background} 

The near horizon geometry of the intersecting branes in~(\ref{conf1})
(excluding the primed, probe branes) is $\text{AdS}_3 \times S^3 \times T^4$.
To keep the connection to the flat space picture manifest we choose to
write the metric and flux in the $\text{AdS}_3\times S^3$ space in Cartesian
and Poincar\'e coordinates, respectively given by\footnote{In order to
go from Cartesian~(\ref{Cartesian}) to Poincar\'e
coordinates~(\ref{Poincare}) one uses the transformations
\begin{equation}
\begin{aligned}
\label{cp}
x_{9} &= R^2 \, u \cos \psi \, , & 
x_{6} &= R^2 \, u \sin \psi \cos \xi \, ,  &
x_{7} &= R^2 \, u \sin \psi \sin \xi \cos \eta \, ,  &
x_{8} &= R^2 \, u \sin \psi \sin \xi \sin \eta \, , \\
x_2   &= l \cos \mu \, , &   x_3 &= l \sin \mu \, .
\end{aligned}
\end{equation}
}
\begin{align}
\label{Cartesian}
ds^2 &= R^2 u^2 \big(-dt^2 + dx_1^2\big) 
  + R_t^2 \, \big( dx_2^2 + \cdots + dx_5^2 \big)
  + {1\over R^2 u^2} \big(dx_6^2 + \cdots + dx_9^2\big)  \,  \\ 
\label{Poincare}
    &\begin{aligned}
     = {R^2 u^2 } \big(-dt^2 + dx_1^2\big) 
        + R_t^2 \,\big(dl^2 + l^2 d\mu^2 &+ dx_4^2 +  dx_5^2\big) 
        +{R^2 du^2 \over u^2} \\
    & +  R^2 \Big(d\psi^2 + \sin^2 \psi \, \big(d\xi^2 +
 \sin^2 \xi d\eta^2\big)\Big)
   \end{aligned}  \\[2ex]
H &= -2R^2 u \, dt \wedge dx_1 \wedge du + 2 R^2 \sin^2 \psi \, \sin\xi \,  d\psi \wedge d\xi \wedge d\eta \, ,
\end{align}
where $u^2= (x_6^2+x_7^2+x_8^2+x_9^2)/R^4$. The dimensionless radius
of $\text{AdS}_3 \times S^3$ is denoted by~$R$ and $R_t$ is the dimensionless
radius of the~$T^4$. These two parameters are independent. In the
remainder of this paper we will always choose to work, for all
metrics, with dimensionless coordinates (so that~$ds^2$ and~$H$ are
also dimensionless). This will be the reason that later, when we take
the Penrose limit of various metrics, we will only scale dimensionless
parameters, while keeping $\alpha'$ fixed.

\subsubsection{The DBI analysis and remarks about calibrations in curved spaces}

Let us consider a $D3$ brane which intersects the $F1-NS5$ system in
flat space as in~(\ref{conf1}) and is placed under the angle~$\alpha$
in the $2-7$ and $3-8$ planes. In Cartesian coordinates its embedding
is given by the following equations
\begin{equation}
\label{simple-cartezian}
x_1=x_4=x_5=x_9=0 \, ,  \quad \quad 
x_2 = \tan \alpha \, x_7 \, , \quad \quad x_3 = \tan \, \alpha x_8 \, .
\end{equation}
Equivalently, if we introduce complex coordinates $\omega = x_7 + i
x_8$ and $z = x_2 + i x_3$ these equations become
\begin{equation}
\label{ansatz-simple}
f(z,\omega) = z - \tan \alpha\,  \omega = 0 \, .
\end{equation}
More generally it is known from K\"ahler calibrations in \emph{flat}
space~\cite{HL,H,Gibbons:1998hm} that all complex submanifolds of
$\mathbb{C}^n$ are volume minimising, and hence solve the D-brane action in the
absence of worldvolume fluxes. In other words, any holomorphic
function $f(z,\omega)= 0$ is a solution to the DBI action in
\emph{flat} space. Different functions have a target space
interpretation of a system of two, three etc.~$Dp/M$ branes which
intersect over a $D_{(p-2)}/M_{(p-2)}$ brane, and are rotated by
arbitrary $SU(2)$ angles.\footnote{For the SU(2) rotations, all branes
are rotated in the same two two-planes.}

We expect that the same result holds in the curved background, namely
if we replace the $F1-NS5$ system with~(\ref{Cartesian}) and treat the
$D3$ brane as a probe in this geometry.  To prove this, let us use the
static gauge in which we identify the worldvolume coordinates
$\sigma^i$ with a subset of target space coordinates as $\sigma^i =
(t,x_6, \omega,\bar{\omega})$ and treat $x_i$~$(i=1,4,5,9)$,~$z$~and
$\bar{z}$ as transverse scalars. We want to check that
\begin{equation}
\label{solu}
{\cal F}_{ij} = 0 \, , \quad z = f(\omega, \bar{\omega}) \,, \quad x_9
= C = \text{const}. \, , \quad x_i = 0 \,\, \quad (i=1,4,5),
\end{equation}
is a solution to the full DBI equations of motion given
in~(\ref{general1}) and~(\ref{general2}), for any holomorphic
function~$f$. Note that we have also put the transverse scalar~$x_9$
to be an \emph{arbitrary} constant since we want to show that,
unless~$C=0$, the above configuration is not a solution to the
near-horizon equations of motion.

Let us first consider the second equation in~(\ref{general2}),
obtained by varying the DBI action with respect to the worldvolume
scalars.  For an arbitrary holomorphic function $f$ and
$x_9=C=\text{const}.$, the $z$ (and~$\bar{z}$) components of this
equation reduce to~$\partial_{\bar{\omega}}\partial_\omega f \equiv
0$. The~$x_9$ component reduces to
\begin{equation}
\label{geom}
{4 x_9 \over u^4 R^2 } = 0 \, , \quad u^2 = x_6^2 + |\omega|^2 + x_9^2 \, .
\end{equation}
We see that $x_9 = \text{const}.\neq 0$ is \emph{not} a solution to the
equations of motion in the near horizon limit. The reason for this can
be seen in flat space: when separating the $D3$ from the $NS5-F1$
system there is an anomalous creation of an $F1$
string~\cite{Hanany:1997ie} which pulls the $D3$ brane toward the
$D1-D5$ system and acts as a source for the worldvolume flux. Hence in
order to obtain the near horizon solution describing a finite~$x_9$
separation one has to turn on the electric field on the brane
worldvolume, as in~\cite{Bachas:2000fr}.  

Note also that although for our configuration the separation in
the direction~$x_9$ is not a free modulus, this is not always the
case. For example for the $D3-D7$ system intersecting over a $D3$
brane $(D3,D7|3+1)$, or a $(D3,D5|2+1)$ system, the separation
(Higgsing) of branes in any common transverse direction can be done
for ``free'': the solution $x^{\text{transverse}}=C=\text{const}.$ is a
supersymmetric solution in flat and $\text{AdS}\times S$
spaces~\cite{Karch:2002sh}. However, note that when taking
the near horizon limit in order to obtain the nontrivial
configuration, one sends not only $u$ to zero but one has to
\emph{simultaneously} scale the constant~$C$ to zero.
This is similar to the situation we will encounter when taking
various Penrose limits in section~\ref{plimt}.

We also have to check that the equation~(\ref{general1}), obtained by
varying the DBI action with respect to the gauge potential,
vanishes. Since there are no RR fields in the target space, the
right-hand side of~(\ref{general1}) is identically zero. The rest is
solved by~(\ref{solu}) with $C=0$, since the antisymmetric part of the
inverse matrix $M^{ij}$ is~\cite{Seiberg:1999vs}
\begin{equation}
\theta^{ij} = \left( {1\over g_{\rm induced} + {\cal F}} \, {\cal F} \, {1\over g_{\rm induced} - {\cal F}}  \right)^{ij}
\end{equation}
i.e.~it vanishes whenever ${\cal F}=0$.

The brane configuration given above can be also described using
calibrations. Following the discussion of calibrated surfaces in
curved spaces~\cite{Gutowski:1999iu,Gutowski:1999tu} we can write the
calibration two-form (calibrating the two-dimensional surface in the
$z,\bar{z}, \omega, \bar{\omega}$ subspace) as
\begin{equation}
\label{clab}
\varphi =  d z \wedge d \bar{z} + {1 \over R^2 \omega \bar{\omega}} d \omega \wedge d \bar{\omega} \, .
\end{equation}
We would like to point out two important points regarding the
construction of solutions to the DBI action in curved spaces using
calibrations.  Firstly, by construction, the calibration method
produces minimal surfaces.\footnote{Here, we are excluding
generalisations of the type presented in
\cite{Gutowski:1999iu,Gutowski:1999tu}. }  Since the DBI action is not
only a geometrical action, establishing the existence of an extremal
point in the truncated part of the full configuration space does not
a-priori imply the existence of an extremal or saddle point in the
full configuration space. Hence, one has to check separately whether
the minimal surface thus constructed is compatible with a truncation
of the full DBI action to its geometric part (i.e.~the Dirac action).

Secondly, the calibration method generically produces minimal surfaces
in \emph{submanifolds} of the full target space. This is basically
dictated by the fact that one needs globally well defined (almost)
complex structures, which generically can only be defined only on
submanifolds (of even dimension). However, the choice of the
submanifold on which we construct the calibration form is not a-priori
restricted. For example, had we chosen the $x_1 =x_4=x_5=0 \, , \, x_9
= \text{const}. \neq 0$ submanifold (as opposed to the submanifold for
which this constant is zero) to construct the calibration
two-form~(\ref{clab}), we would have ended up with a minimal
two-surface on this submanifold. This one does not, however, lift to
an extremal two-surface in the full space, as one can see from
equation~(\ref{geom}).\footnote{Note that the $x_9 = 0$ surface
actually lifts to a \emph{maximal} two sphere in the full space. This
is just an illustration of the fact that a minimum in a subspace of
the full configuration space can lift to a saddle point in the full
configuration space.} Hence, again, one needs to make sure that the
truncation which one makes when restricting to a submanifold is a
consistent one.


\subsubsection{The geometry of the diagonal brane}

Let us now try to understand the geometry of a \emph{generic} diagonal
$D3$ brane. In global coordinates (for definitions of various
coordinate systems see~(\ref{cp}) as well as~(\ref{AdS-global})
and~(\ref{standard-s3}) in the appendix) a generic diagonal $D3$ is
given by the equations
\begin{equation}
\label{glob-embd}
\begin{aligned}
\psi &= {\pi \over 2}\, , & \sigma &= 0 \, , \quad x_4 = x_5 =0 \,, \\
f(z,\omega) &= 0 \, ,     & \omega &= R^2 \sin\xi \big(\cosh\rho \cos
\tau - \sinh \rho\big)\, e^{i\eta}\,   \, .
\end{aligned}
\end{equation}
As before $f$~is a holomorphic function of $z$ and $\omega$.  To
understand the geometry of the brane, let us first understand the
simple example of the holomorphic function
\begin{equation}
\label{ansatz1}
f(z,\omega) = (z - \tan \alpha \, \omega)\omega - c \, .
\end{equation}
We see that there are two asymptotic solutions to the equation~$f(z, \omega)=0$,
\begin{eqnarray}
\label{large}
|\omega| \ll |c| && \quad |z| \rightarrow \infty \, , \\
\label{small}
|\omega| \gg |c| && \quad z \rightarrow \tan \alpha \, \omega \, .
\end{eqnarray}
These correspond to two different $D3$ branes: the first one is
embedded along the $T^2(z)$ torus in $T^4$ and the second one is
diagonally embedded between the $S^2 \times \rho$ part in $\text{AdS}_3
\times S^3$ and the $T^2(z)$ torus in $T^4$. So, similar to the
situation in flat space, the function~(\ref{ansatz1}) describes two $D3$
branes that intersect under the angle ${\pi \over 2} - \alpha$ in
the $\text{AdS}_3\times S^3\times T^4$ space.

To see this in more detail, let $\alpha =0$ (this is the case of two
orthogonally intersecting branes) and focus on the region near the
boundary, $\rho \rightarrow \infty$. Then we have
\begin{equation}
|\omega| =  \tfrac{1}{2} R^2 \,  e^{\rho} \, \big|\sin \xi\, (\cos \tau - 1)\big| \, .
\end{equation}
We see that for $\xi \neq 0$ and $\tau \neq 0 \, (\text{mod} \,\, \, 2
\pi)$, $|\omega|$ is large, so we recover the asymptotic solution
(\ref{small}) which is the $\text{AdS}_2\times S^2$ brane. As we approach
the origin of $\text{AdS}_3$ ($\rho \rightarrow 0$), $|\omega|$ becomes
\begin{equation}
|\omega| = R^2 \big|\sin \xi (\cos \tau - \rho)\big| \leq R^2 \, .
\end{equation}
So by choosing the constant $c$ in~(\ref{ansatz1}) larger than~$R^2$ we
see that~(\ref{large}) holds, hence the brane asymptotes to the~$\text{AdS}_2
\times T^2(z)$ brane.  In summary, as we move in the radial direction
toward the center of AdS the brane system interpolates between
the~$\text{AdS}_2 \times S^2$ brane and the~$\text{AdS}_2 \times T^2$ brane.
If~$\alpha\neq 0$ then the asymptotic region of large~$\rho$
corresponds to a \emph{single} non-conformal
brane~(\ref{ansatz-simple}), which nontrivially mixes the~AdS,
sphere and torus parts, while the~$\rho \rightarrow 0$ region remains
unchanged.

These kind of interpolating brane solutions are especially interesting
in the light of the defect (conformal) field theory/AdS
correspondence. Particularly interesting examples are those of $D5$
and $D7$ branes in $\text{AdS}_5 \times S^5$ which deserve a separate
study~\cite{sarkissian}. These embeddings of branes are supposed to
correspond to an RG flow between two different defect conformal field
theories.

Finally, let us conclude this section with a brief comment on
generalisations of the function $f$ in~(\ref{solu}) to higher degree
(holomorphic) polynomials. These generalisations lead to
configurations of multiply-intersecting brane systems, all rotated in
the same two two-planes $2-7$ and $3-8$ but with different angles. In
order to describe~$SU(3)$ and~$SU(4)$ rotations (in three and four
two-planes) one needs to excite four~$(z_1,z_2)$ and
six~($z_1,z_2,z_3$) transverse scalars respectively. The results are
the similar to those that we have presented for the~$SU(2)$ case.


\subsubsection{CFT symmetries of the diagonal branes}

In the context of CFT constructions of D-branes one usually starts
from a particular set of conditions imposed on the CFT currents, and
then one deduces the effective geometry associated to these
D-branes. Here, we are facing an opposite problem. The D-branes which
were constructed in the previous sections are given by their
effective, geometric embeddings. We would now like to deduce something
about the CFT symmetries that they preserve.

Our strategy will be the following. We will first determine the
\emph{geometric} isometries preserved by the effective D-brane
hypersurface.  We will then try to argue how these geometrical
isometries get lifted to worldsheet symmetries.\footnote{The
geometrical isometries lead to conserved currents on the worldline of
a point particle. In that sense the effective geometry is the geometry
as seen by a point particle.}  This approach, however, cannot directly
tell us about the conserved charges that are stringy in nature
(winding charges, for example).

We will restrict our analysis to the simplest case of the diagonal
$D3$ brane~(\ref{simple-cartezian}).  In the AdS-Poincar\'e
coordinates and cylindrical coordinates in the $2-3$ plane, as given
in~(\ref{Poincare}), this brane is described by an effective embedding
equation
\begin{equation}
\label{eqns-sym}
x_2^2 + x_3^2 = \alpha^2 u^2  \sin^2 \xi \, , \quad  {x_2 \over x_3} = \tan \eta  \, , \quad  \psi = {\pi \over 2} \, , \quad x_1 = x_4 = x_5 = 0\,  .
\end{equation}
We see that this surface nontrivially mixes the $\text{AdS}_2$, $S^2$ and
$T^2$ submanifolds of the~$\text{AdS}_3 \times S^3 \times T^4$ background.
As a first step in the analysis we need to determine the isometries
preserved by the~$\text{AdS}_2$ submanifold in~$\text{AdS}_3$ and the~$S^2$
submanifold in~$S^3$.

An $\text{AdS}_2$ submanifold in $\text{AdS}_3$ is defined by a twined conjugacy
class
\begin{eqnarray}  
\label{AdS2}
{\cal C}_{(\omega,g)} = \bigg{\{} \omega(h) g  h^{-1} | \, \, \forall
h  \in SL(2,R) \, \, \bigg{\}} \, 
\quad\text{with}\quad
\omega(h) = \omega_0^{-1} h \omega_0  \,,\quad \omega_0 = \begin{pmatrix} 0 & 1 \\
1 & 0 
\end{pmatrix}\,.
\end{eqnarray}
Here $\omega(h)$ is an outer automorphism of the group. Note that all
elements in the same conjugacy class have a fixed value of the trace,
equal to~$\tr(\omega_0 g)$.  The $\text{AdS}_3$ metric is invariant under 
\emph{separate} left and right group multiplications. These are
related to two copies of the $SL(2,R)$ currents in the bulk CFT.
The submanifold~(\ref{AdS2}) (i.e.~the constant trace) is invariant
only under the \emph{simultaneous} right and twisted left
multiplications
\begin{equation}
\label{AdS2sym}
\tr\big(\omega_0 g\big) = \tr\Big(\omega_0 (\omega_0^{-1}  h \omega_0) g h^{-1}\Big)   \, .
\end{equation}
These isometries are related to the following set of worldsheet
currents (on an open string which is restricted to move on the~$\text{AdS}_2$
surface)
\begin{equation}
J =\omega_0 \bar{J} \omega_0 \, .
\end{equation}
However, our brane is a subsurface in $\text{AdS}_2$, hence it breaks the
symmetries further. To determine the residual isometry we will need
the transformation properties of the~$\text{AdS}_2$ coordinates
under~(\ref{AdS2sym}).  This can easily be determined using the
parametrisation~(\ref{SL2R}) and~(\ref{parm}).  The coordinate~$u$
transforms as
\begin{equation} 
u ' = D^2 u -2 {C \, D\,\over R} u \, t - \left({C\over R}\right)^2 \left({1\over u} + u (-t^2 +
x_1^2)\right) \,, 
\end{equation} 
where $A,B,C$ and~$D$ are parameters of an arbitrary~$SL(2,R)$
matrix. Note that for our
brane~(\ref{simple-cartezian}) the \text{AdS}~coordinate~$x_1$ is zero. We see
that under the action of the matrix
\begin{equation}
\label{u-const}
h = \begin{pmatrix} 1 & B \\
                    0 & 1 \end{pmatrix} = e ^{{B \over 2} (\sigma^{+})} \, , \quad  \sigma^{+} = \sigma^1 + i \sigma^2 \, ,
\end{equation}
the coordinate~$u$ does not change.

For the isometries involving a sphere part of the configuration, note
that an arbitrary two sphere~$S^2$ in an~$SU(2)$ manifold is given by
the conjugacy class
\begin{equation}
{\cal C}_{g} =  \bigg{\{} h g h^{-1} | \, \forall h \in SU(2) \bigg{\}} \, . 
\end{equation}
It is invariant under the adjoint action~$g \rightarrow h^{-1} g
h$. This implies that the~$SU(2)$ currents~$J_{SU(2)}, \bar{J}_{SU(2)}$
preserved by the~$S^2$ brane are
\begin{equation} 
J_{SU(2)} =  \bar{J}_{SU(2)}  \, .
\end{equation}
For the isometries of our brane we will also need the transformation
properties of the coordinate~$\xi$ under the adjoint action~$(g
\rightarrow h^{-1} g h)$,\footnote{Since our brane mixes only the $S^2$
given by $\psi = \pi/2$ with the $\text{AdS}_2$ factor, we do not need a
generic transformation, but we restrict here to the big~$S^2$ in~$S^3$.}
\begin{equation}
\cos \xi' = (a a^* - b b^*) \cos \xi  - \Real(a b^*) \sin \xi \cos \eta - \Imag(a b^*) \sin \xi \sin \eta \, .
\end{equation}  
Here $a,b$ are the complex parameters of an $SU(2)$
matrix.\footnote{We write a generic~$SU(2)$ matrix as 
\begin{equation}
\begin{pmatrix}
a & b \\
-b^* & a^*
\end{pmatrix}
\, , \quad |a|^2 + |b|^2 = 1 \, .
\end{equation}}
We see that $\xi$ is unchanged if and only if~$b=0$. This
transformation is just a~$U(1)_{\eta}$ symmetry: it acts on~$\eta$
as~$\eta \rightarrow \eta + \alpha$ (where $\alpha=\text{const}.$ is a
parameter of the transformation) while keeping~$\xi$ and~$\psi$
unchanged.

Next, we need to determine which of the above symmetry transformations
leave the conditions~(\ref{eqns-sym}) unchanged. The first condition
is invariant only under the transformations which separately preserve
the~$u$, $\xi$, $x_2$ and~$x_3$ coordinates: these are
the~$U(1)_{\sigma^+}$ and~$U(1)_{\eta}$ rotations. The second condition
however breaks the~$U(1)_{\eta}$ isometry, implying that the only
preserved current which originates from the~$\text{AdS}_3$ and~$S^3$
isometries is
\begin{equation} 
J_{+} =\omega_0 \bar{J}_+ \omega_0 \, .
\end{equation} 
As far as the isometries of the torus are concerned, the
conditions~(\ref{eqns-sym}) obviously break the~$U(1)_{x_i}$
isometries (with $i=2,3,4,5$). However, since the directions~$x_4$
and~$x_5$ are Dirichlet, we know that the (stringy) winding
currents
\begin{equation} 
J_T^{4} = - \bar{J}_T^4 \ , \quad J_T^{5} = -\bar{J}_T^5 \,,
\end{equation}
are preserved. On the other hand, the conditions on~$x_2$ and~$x_3$
are neither Dirichlet nor Neumann conditions, and to determine the
(stringy) currents associated to the first two conditions
in~(\ref{eqns-sym}), one needs to analyse the full sigma model subject
to these boundary conditions. We will not perform this analysis here.

\subsection{Diagonal $D4$ brane in the $\text{AdS}_3 \times S^3 \times S^3
  \times R$ background and its symmetries} 

Another interesting configuration of a diagonal D-brane can be
obtained from the configuration~(\ref{conf2}) of intersecting
branes. As before, we replace the first three D-branes
in~(\ref{conf2}) by the near-horizon limit of the corresponding
supergravity solution and treat the~$D4$ brane as a probe. The probe
can interpolate between the two reference positions marked
in~(\ref{conf2}).

The metric of the near-horizon limit of the supergravity solution for
the bound state of the~$N_F^{(1)}$~$NS5_1$ branes,
the~$N_F^{(2)}$~$NS5_2$ branes and the~$N_{T}$ fundamental strings is
\cite{Cowdall:1998bu}
\begin{equation}
\label{carmer}
ds^2 = {r^2r'^2\over N_T}\Big(-dt^2+dx_1^2\Big)+{N_F^{(2)}\over r^2}\Big(dx_2^2+dx_3^2
+dx_4^2+dx_5^2\Big)+{N_F^{(1)}\over r'^2}\Big(dx_6^2+dx_7^2
+dx_8^2+dx_9^2\Big) \,,
\end{equation}
where $r^2=x_2^2+x_3^2+x_4^2+x_5^2$ and
$r'^2=x_6^2+x_7^2+x_8^2+x_9^2$. As before, we have factored out the
dimensionful parameter~$\alpha'$ so that all coordinates and all
parameters are dimensionless.  It is convenient to introduce new
variables \cite{Gauntlett:1998kc,deBoer:1999rh}
\begin{equation}
\label{desf}
u={rr'\over N_T^{1/2}R}\, ,\;\;\;
\lambda=R\left(\sqrt{{N_F^{(1)}\over N_F^{(2)} }}\, {\rm log} \, r-
\sqrt{{N_F^{(2)}\over N_F^{(1)} }}\, {\rm log} \, r'\right)\, , \;\;\; 
{1\over R^2}={1\over N_F^{(1)}}
+{1\over N_F^{(2)}} \, .
\end{equation}
In these variables the metric~(\ref{carmer}) and the accompanying NS
3-form can be written as
\begin{eqnarray}
\label{metr}
ds^2 &=& R^2\left(u^2(-dt^2+dx_1^2)+{du^2\over u^2}\right)+
d\lambda^2+N_F^{(2)}d\Omega_{(1)}^2+
N_F^{(1)}d\Omega_{(2)}^2  \, \\
\label{volf}
H &=& 2\, R^{-1}\epsilon_{\text{AdS}_3}+2\, N_F^{(2)}\epsilon_{S^{3(1)}}
+2\, N_F^{(1)}\epsilon_{S^{3(2)}} .
\end{eqnarray}
Here $d\Omega_{(1)}^2$ and $\epsilon_{S^{3(1)}}$ are the metric and
volume forms for the unit three-sphere in the space spanned by the
$x_2,x_3,x_4,x_5$ coordinates. Similarly, $d\Omega_{(2)}^2$ and
$\epsilon_{S^{3(2)}}$ are the metric and volume form for the unit
three-sphere in the space spanned by the $x_6,x_7,x_8,x_9$
coordinates.  We see that~(\ref{metr}) and~(\ref{volf}) manifestly
exhibit the $\text{AdS}_3\times S^{3}\times S^3\times S^1$ structure of the
near-horizon geometry.

To find the effective geometry for the diagonal $D4$ probe brane in
this background we follow the same logic as in the previous
section. As before, it can easily be shown that the embedding given by
\begin{equation}
\label{holbr}
x_1=x_5=x_9=0\, , \quad  x_7+ix_8=f(x_2+ix_3) \, , \quad {\cal F}=B+F = 0 \, ,
\end{equation}
with $f$ an arbitrary holomorphic function, solves the DBI action. This
is in agreement with the general statements about calibrated surfaces
in curved spacetime.  The calibration two-form (calibrating the
$x_2,x_3,x_7,x_8$ part of the D-brane surface) is
\begin{equation}
\varphi = {1\over x_2^2 + x_3^2} dx_2 \wedge dx_3 + {1\over x_7^2 + x_8^2} dx_7 \wedge dx_8 \, .
\end{equation}

To say something about the symmetries and currents preserved by this type
of brane we restrict to the simpler case of a brane with embedding
\begin{equation}
\label{eqint}
x_1=x_5=x_9=0\, , \quad x_2= \tan \alpha x_7\, , \quad x_3= \tan \alpha x_8\, .
\end{equation}
Using the coordinates~(\ref{standard-s3}) and~(\ref{desf})  this
surface can be written as
\begin{equation}
\begin{aligned}
\label{glintsph}
x_1&=0\, , \quad \psi_1= \psi_2= {\pi\over 2} \, , \quad
\eta_1=\eta_2\, ,  \\
\left( {r\over r'} \right)^{a + {1\over a}} &= \left(\tan \alpha{
\sin\xi_2\over \sin\xi_1}\right)^{a+{1\over a}}=
(R\sqrt{N_T})^{{1\over a}-a}
e^{{2\lambda\over R}} u^{{1\over a}-a} \, ,
\end{aligned}
\end{equation}
where $a=\sqrt{N_F^{(1)}/N_F^{(2)}}$. We see that for generic values
of the parameters, the $D4$ brane mixes in a nontrivial way all four
submanifolds. However, when $a=1$ the $u$-dependence
in~(\ref{glintsph}) drops and one is left with a brane that has the
geometry of an $\text{AdS}_2$ brane times a $3$-brane in the $S^3\times
S^3\times S^1$ subspace. This brane is defined by the equations
\begin{equation}
\label{dthb}
\psi_1=\psi_2={\pi\over 2}\, , \quad
\eta_1=\eta_2\, ,\quad e^{{\lambda\over R}}= \tan \alpha \,{
\sin\xi_2\over \sin\xi_1}\, .
\end{equation}
It is easy to find the symmetries preserved by this $D4$ brane. It
trivially inherits those of the $\text{AdS}_2$ brane. For the
three-dimensional part wrapping the spheres, we note that the first
and third condition in~(\ref{dthb}) are preserved by separate $U(1)$
rotations in the directions $\eta_1$ and $\eta_2$.  Preservation of
the second condition, however, relates these currents as 
\begin{equation}
\label{consym}
K_3+\bar{K}_3=L_3+\bar{L}_3\, , 
\end{equation}
where $K_3$,${\bar K}_3$ and $L_3$,${\bar L}_3$ are the left/right
currents corresponding to $U(1)_{\eta_1}$ and $U(1)_{\eta_2}$
symmetries.

Another interesting brane can be derived by taking the holomorphic
function~$f$ in~(\ref{holbr}) to be $z=\beta/w$.  Using the
coordinates~(\ref{standard-s3}) and the definition of $u$
in~(\ref{desf}) we get a $D4$ brane with the embedding\footnote{These
formulae are valid only when $\beta\neq 0$.}
\begin{equation}
\label{nholb}
 x_1=0\, ,\quad
\psi_1=\psi_2={\pi\over 2}\, , \quad
\eta_1=-\eta_2\, ,\quad u={\beta \over R\sqrt{N_T}}{1
\over \sin\xi_1\sin\xi_2}.
\end{equation}
Now the geometry of the D4-brane is a product of a D3-brane on
$\text{AdS}_3\times SU(2)\times SU(2)$ defined by~(\ref{nholb}) and a~U(1)
circle wrapped by~$\lambda$.

\section{Turning on a non-diagonal worldvolume flux}
\label{seven}

Up to now, our analysis has been restricted to situations where~${\cal
F} = 0$. The mixing between the subspaces of the target space manifold
was geometric: the brane worldvolumes were embedded in the target
space in a diagonal way between the two submanifolds.  We will now
show that there is yet another way of implementing a mixing between
target space submanifolds. Namely, for branes whose worldvolume is a
product~$\text{AdS} \times S$ it is possible to turn on a worldvolume flux
which mixes the~AdS and~$S$ directions.  Moreover, unlike in the
other known cases of branes with fluxes in~$\text{AdS}\times S$
spaces~\cite{Bachas:2000fr}, the presence of the non-diagonal flux in
these cases does not modify the brane
geometry.\footnote{In~\cite{Bachas:2000fr} a configuration of an
$\text{AdS}_2 \times S^2$ brane with electric and magnetic worldvolume fluxes
has been considered. It was shown that one can turn on
electric/magnetic fields which are \emph{each separately} proportional
to the volume forms of the $\text{AdS}_2/S^2$ parts, and still preserve
supersymmetry. The effect of the magnetic flux on the geometry of the
embedded brane was that it changed (decreased) the size of the $S^2$
sphere on which the brane was wrapped. Similarly, the electric flux
led to an asymptotically $\text{AdS}_2$ manifold, whose curvature was larger
than that of the background $\text{AdS}_3$ space. The sources of the
magnetic/electric fluxes were $(p,q)$ strings connecting a $D3$ and a
background system, along direction $9$ in~(\ref{conf1}). Since a
$(p,q)$ string pulls a $D3$ branes, its deforms its surface, and taking
the near horizon limit we focus on a ``deformed'' part of a $D3$ brane,
which leads to these kinds of geometries.}

In a flat space it is well known that performing a T-duality
transformation under an angle leads to brane configurations with
non-vanishing fluxes. Hence using this property it is possible to
``trade'' non-orthogonally intersecting brane systems for orthogonally
intersecting ones with flux. We want to show that a similar logic goes
through in a curved target space.

Our starting point is the supersymmetric flat space configuration,
\begin{equation}
\label{conf3}
\begin{array}{lcccccccccc}
D3   & 0 & - & 2 & 3 & - & - & - & - & 8 & -  \\ 
D5'  & 0 & - & 2 & - & 4 & 5 & 6 & - & 8 & -  \\ 
D5'' & 0 & 1 & - & - & 4 & 5 & 6 & 7 & - & -  \\
\end{array}
\end{equation}
We focus on the $D5$ brane which in a supersymmetric way intersects the
``reference''~$D5'$ and~$D5''$ branes under an angle~$\alpha$,
\begin{equation}
x_2 = \tan \alpha \, x_1\, , \quad x_8 = \tan \alpha \, x_7 \, .
\end{equation}
It extends in directions $4,5$ and $6$. Applying a T-duality
transformation in the directions~$1$ and~$8$ leads to the following
configuration:
\begin{equation}
\label{conf4}
\begin{array}{lcccccccccc}
D3   & 0 & 1 & 2 & 3 & - & - & - & - & - & -  \\ 
D5'  & 0 & 1 & 2 & - & 4 & 5 & 6 & - & - & -  \\ 
D5'' & 0 & - & - & - & 4 & 5 & 6 & 7 & 8 & -  \\
D7   & 0 & 1 & 2 & - & 4 & 5 & 6 & 7 & 8 & - 
\end{array}
\end{equation}
The $D7$ brane originates from the diagonal~$D5$ brane. Note that
the~$D7$ carries nontrivial worldvolume flux
\begin{equation}
\label{fflux}
F_{12} = - \cot \alpha \, , \quad \quad F_{78} = \tan \alpha \, , 
\end{equation}
due to the initial diagonal embedding. Note also that, in the absence
of fluxes, the configuration~(\ref{conf4}) is
\emph{non-supersymmetric} and hence no-longer T-dual to~(\ref{conf3}).

The $D7$ brane orthogonally intersecting the~$D3$ brane leads, in the
near horizon limit, to an~$\text{AdS}_4 \times S^4$ brane. This~$D7$
brane-embedding solves the DBI equations of motion, though it is
obviously not a supersymmetric configuration. This was explicitly
checked in \cite{Skenderis:2002vf} by performing a kappa symmetry
analysis.\footnote{One might think that taking the near horizon limit
of a non-supersymmetric configuration might lead to a supersymmetric
configuration. However, in the Penrose limit the $\text{AdS}_4 \times S^4$
brane reduces to a non-supersymmetric $(+,-,3,3)$ brane, and taking a
Penrose limit cannot decrease the amount of supersymmetry.}  We will
now show that the \emph{flat} space flux~(\ref{fflux}) solves the
equations of motion in an $\text{AdS}_5\times S^5$ background.  Moreover, due
to the supersymmetric flat space origin of this flux, it is clear that
this configuration is supersymmetric. Note that the second component
of the worldvolume flux~(\ref{fflux}) couples the $\text{AdS}_5$ and $S^5$
components of the metric. To see this explicitly we rewrite the flux
using Poincar\'e coordinates in the $\text{AdS}_5$ part and using the
following parametrisation for the five sphere:
\begin{equation}
\label{phic}
x_7 = u \cos \theta \cos \varphi \, , \quad 
x_8 = u \cos \theta \sin \varphi \,,  \quad 
x_i  = u \sin \theta \,\Omega_i \,,
\end{equation}
with $i=4,5,6,9$ and $\sum_i\Omega_i^2 = 1$.  In these coordinates the
metric on the five-sphere becomes
\begin{equation}
\label{S5}
ds^2_{S^5} = \cos^2 \theta d\varphi^2  + d \theta^2 + \sin^2 \theta \Big( d\psi^2 + \sin^2 \psi \big(d \xi^2 + \sin^2 \xi d \eta^2\big)\Big) \, ,
\end{equation}
and the flux~(\ref{fflux}) is given by
\begin{equation}
\label{flux-all}
F = -\cot \alpha \, dx_1 \wedge dx_2 + \tan \alpha \, R^4 u  \cos^2 \theta \, du \wedge d \varphi - \tan \alpha \, R^4 u^2 \sin \theta \cos \theta d \theta \wedge d \varphi \, .
\end{equation}

Let us now show that the configuration given above solves the DBI
action for the $D7$ probe in $\text{AdS}_5\times S^5$. The calculation
simplifies greatly in Cartesian coordinates, in which the metric is
given by
\begin{equation}
ds^2 =  R^2 u^2 \Big( - dx_0^2 + dx_1^2 + dx_2^2 + dx_3^2 \Big) + {1
  \over R^2 u^2} \Big( dx_4^2+ \cdots + dx_9^2 \Big) \,,
\end{equation}
with $u^2 = (x_4^2 + \cdots + x_9^2)/R^4$.  Further simplifications
are obtained by going to the static gauge,
\begin{equation}
\sigma_i = x_i \,  , \quad \text{where}\quad i=0,1,2,4,5,6,7,8 \, .
\end{equation}
By examining the Wess-Zumino term it is clear that the $D7$ brane with
the above embedding does not couple to the background RR-flux.  In
this case the general DBI equations of motion given in
(\ref{general1}) and (\ref{general2}) reduce to
\begin{equation}
\label{extr}
{\cal K}_{{\mu}} = 0 \, , \quad  \partial_{i} (\sqrt{- \det M} \theta^{ij}) = 0 \,,
\end{equation}
where $M_{ij}$, $G^{ij}$ and $\theta^{ij}$ are defined
in~(\ref{MT}). In the absence of (covariant) flux~${\cal F}$ the
quantity ${\cal K}_{\mu}$ reduces to the trace of the second
fundamental form. In our case of a maximal $\text{AdS}_4$ and a maximal
$S^4$~(given by $x_3 =x_9 = 0$) the second fundamental form vanishes
identically. This simplifies the calculation of ${\cal K}_{\mu}$ in the
presence of flux. It is easy to see that the non-geometric terms
(i.e.~flux-dependent terms) in ${\cal K}$ cancel separately. One uses
\begin{equation}
\begin{aligned}
\det M &= - {1\over u^8R^8 \tan^2 \alpha} \Big( 1+ u^4R^4 \tan^2
\alpha\Big)^2  \,, \\[1ex]
G^{00} &= -{1 \over u^2R^2} \, , &
G^{44} &= G^{55} = G^{66} = u^2R^2\,,\\[1ex]
G^{11} &= G^{22} = {u^2 R^2\tan^2\alpha \over  1+u^4R^4\tan^2\alpha}\,, &
G^{77} &= G^{88} = {u^2R^2 \over 1+u^4R^4 \tan^2\alpha}\,,
\end{aligned}
\end{equation}
with all other components vanishing.  The second equation
in~(\ref{extr}) vanishes since~$\theta^{ij}$ reduces to
\begin{equation}
\theta^{12} = {\tan \alpha \over  1+u^4R^4\tan^2 \alpha } \, , \quad \theta^{78} = {-  u^4R^4 \tan \alpha \over 1+u^4R^4\tan^2\alpha } \,,
\end{equation} 
with vanishing other components for the flux (\ref{fflux}).

Finally, we would like to point out that this kind of D-brane, with
non-diagonal flux, is not covered by the general analysis
of~(generalised) calibrations with worldvolume fluxes as studied
in~\cite{Marino:1999af}. The cases covered in~\cite{Marino:1999af} are
supersymmetric in the absence of worldvolume flux, while our D-brane
is non-supersymmetric if the flux is zero. The non-diagonal flux will
also lead to an interesting Penrose limit which we will discuss in
section~\ref{pflux}.

\section{Penrose limits}
\subsection{Penrose limit of the diagonal $D3$ brane: a brane with a pulse}
\label{plimt}
It is usually said that in order to have a nontrivial Penrose limit of
a brane in some background, one needs to take the limit along a
geodesic which belongs to the brane. This statement is intuitively
understandable: in the Penrose limit an infinitesimal region around
the geodesics gets zoomed out. Hence, those parts of the brane which
are placed at some nonzero distance from the geodesic get pushed off
to infinity.  However, this reasoning can be circumvented if the
distance between the geodesic and the brane is determined by free
parameters of the solution. In that case one can take the Penrose
limit along a geodesic that \emph{does not} belong to the brane, as
long as the parameter labeling the brane in a family of solutions is
appropriately scaled.

For example, let us consider the family of solutions corresponding to
two intersecting D-branes and let us take the Penrose geodesic to lie
on one of the two branes. Then the Penrose limit of the other brane
can be nontrivial if, while taking the Penrose limit of the target
space metric, we simultaneously scale the angle between the two branes
to zero.  It should be emphasised that the final configuration
obtained in this way is different from the one which is obtained by
first sending the angle to zero and then taking the Penrose limit of
the metric.

To see how these ideas work in practice, we consider in the next two
sections the Penrose limit of the family of diagonal $D3$ branes given
in~(\ref{simple-cartezian}) with the Penrose geodesic taken to be a
generic null geodesic that mixes the AdS, sphere and torus parts and
does not belong to the brane worldvolume. In the first
section,~\ref{ses1}, we work out the Penrose limit of the background,
obtained by using a generic geodesic.  In the second
section,~\ref{ses2}, we restrict the analysis to a subclass of
geodesics that do not wind on the torus (and do not belong to the
brane worldvolume) and apply a Penrose limit on the brane while
simultaneous taking the angle to zero. We end up with a brane which is
the same as the one which we would get if we would first have sent the
angle to zero, except that there is now a null pulse propagating on
the brane worldvolume.

\subsubsection{Penrose limit of the $\text{AdS}_3 \times S^3 \times T^4$ background}
\label{ses1}

Let us consider a null geodesic which mixes the AdS, sphere, and torus
parts, extends in the subspace $(t, u, \psi, x_2, x_3)$ in Poincare
coordinates~(\ref{Poincare}) and is placed at a constant value of
transverse coordinates\footnote{The analysis of a geodesic which also
involves the direction $\xi$ is more involved but leads to a similar
conclusion, namely that for the generic
brane~(\ref{simple-cartezian}) there is no null geodesics which
belongs to its worldvolume. The geodesics we consider here are slight
generalisations of the geodesics analysed in~\cite{Blau:2002mw}.}
\begin{equation}
\label{ged0}
x_1 = x_4 = x_5 = 0 \, , \quad \xi= {\pi\over 2} \, , \quad  \eta = 0
\, .
\end{equation}
As far as the Penrose limit of the metric goes, these values of the
transverse coordinates are irrelevant. The choice which we have made
here is such that the analysis of the Penrose limit of the brane in
the next section is simplified.  To find the geodesic we make use of
the following conserved quantities (dots denote derivatives with
respect to the affine parameter~$\tau$):
\begin{equation}
R^2 u^2 {\dot t} = E \, , \quad R^2 {\dot \psi} = l \, , \quad R_t^2
{\dot x_2} = p_2 \, , \quad R_t^2 {\dot x_3} = p_3 \, .
\end{equation}
We also use the null condition $ds^2 = 0$. The geodesic is given by
\begin{equation}
\label{ged1}
\begin{aligned}
u &= {E \over L} \sin\left({L\over R^2} \tau\right) \, , \quad t = - {L\over E} \, \cot\left({L \over R^2} \tau\right) \, , \quad \psi = {l \over R^2} \tau \,,  \\
x_2 &= {p_2 \over R_t^2} \tau \, , \quad x_3 = {p_3 \over R_t^2} \tau \, , \quad L^2 \equiv l^2 + {R^2 \over R_t^2} \big(p_2^2 + p_3^2\big) \, .
\end{aligned}
\end{equation}
Next we go to an adapted coordinate system in which one of the new
coordinates~(namely the coordinate~$\tilde{u}$) is identified with the
affine parameter along the null geodesic~(i.e.~$\tilde{u} = \tau$).
All other coordinates are chosen in such a way that the geodesic
is located at the origin in the remaining directions. More precisely,
we introduce new coordinates
\begin{equation}
\label{newc}
(\tilde{u}, \tilde{v}, \tilde{\phi}, \tilde{\xi}, \tilde{\eta}, \tilde{x_i})
 \,,\quad\text{with}\quad i=1,2,3,4,5 \, .
\end{equation}
We require that in the new coordinates the metric components
$g_{\tilde{u}\tilde{u}} = g_{\tilde{u}\tilde{x}_2} =g_{\tilde{u}
\tilde{x}_3} = 0$ and $g_{\tilde{u}\tilde{v}}=R^2$. To find a set of
coordinates which fulfills these requirements we use the ansatz
\begin{equation}
u =  {E \over L} \, \sin\left({L \over R^2} \tilde{u}\right) + f_1(\tilde{v},
\tilde{\phi}, \tilde{\xi}, \tilde{\eta}, \tilde{x_i})\, ,  \quad \psi
= {l \over R^2} \, \tilde{u} + f_2 (\tilde{v}, \tilde{\phi},
\tilde{\xi}, \tilde{\eta}, \tilde{x}_i) \, ,\quad\text{etc.}
\end{equation}
One possible choice for an adapted
coordinate system with these properties is given in the appendix, see
equation~(\ref{adapt}).  By writing the metric in this adapted
coordinate system (see~(\ref{rose1})) one can see that in order to
obtain a finite result for the metric in the $\lambda \rightarrow 0$
limit, we have to scale the tilded coordinates and the parameters~$R$
and~$R_t$ according to
\begin{equation}
\label{scaling}
\begin{aligned}
R   &\rightarrow \lambda^{-1} R\,, &
R_t &\rightarrow \lambda^{-1} R_t \,, &   
\tilde{u} &\rightarrow \lambda^{-2} \tilde{u} \,, & 
\tilde{v} &\rightarrow \lambda^4 \tilde{v} \,,\\[1ex]
\tilde{\phi} &\rightarrow \lambda \,\tilde{\phi} \,, &
\tilde{\xi}  &\rightarrow \lambda \tilde{\xi} \,, & 
\tilde{\eta} &\rightarrow \lambda \tilde{\eta} \,, &
\tilde{x}_i  &\rightarrow \lambda \tilde{x}_i \,,\quad\text{with}\quad i=1,2,3,4,5 \, .
\end{aligned}
\end{equation}
Recall that we have chosen to work with dimensionless coordinates;
hence we take a Penrose limit by scaling only the dimensionless
quantities and keeping~$\alpha'$ fixed.
In the limit (\ref{scaling}) the metric (\ref{rose1}) in adapted
coordinates reduces to metric of the following form (expressed in
\emph{Rosen coordinates}):
\begin{equation}
\label{rose}
ds^2  =  2 R^2 d\tilde{u} d\tilde{v} + \sum_{ij} C_{ij}(\tilde{u})\,dy^i dy^j
\end{equation}
where $y^i$ denotes all coordinates apart from $\tilde{u}$ and
$\tilde{v}$, and $C_{ij}$ is a symmetric \emph{non-diagonal}
matrix. To put this metric into more familiar Brinkman coordinates one
needs to follow a procedure outlined in~\cite{Blau:2002mw}.  For us
however, the explicit form of the general metric~(\ref{rose1}) in
Brinkman coordinates will not be relevant, so we do not perform these
coordinate transformations here.\footnote{For generic values of
the momenta~$p_2$ and~$p_3$ one is not automatically guaranteed that
the metric~(\ref{rose}) describes a Lorenzian symmetric~(Cahen-Wallach)
space, i.e.~that it is one of the Hpp-waves. For that one needs to
check that the matrix~$A_{ij}$ in the metric written in Brinkman
coordinates,
\begin{equation}
ds^2 = 2 dx^{+} dx^{-} + \sum_{ij} A_{ij}(x^{-})x^i x^j (dx^{-})^2 + dx_i^2 
\end{equation}
is a constant matrix. In our case a change of coordinates from Rosen to
Brinkman is more involved due to the fact that~$C_{ij}$ is
non-diagonal. }

On the other hand, one can check that in the $p_2 = p_3 = 0$ case,
which will be used in the next section, equation~(\ref{rose1}) reduces
to a standard Hpp-wave in Brinkman coordinates, after the change of
coordinates given by~(\ref{change}).  The wave metric is given by
\begin{multline}
\label{limi}
ds^2 = 2 dx^+ dx^- - {4L^2 \over R^4} \Big(dy_1^2 + dy_2^2 + dz_1^2 + dz_2^2\Big)
\big(dx^+\big)^2 + dy_1^2 + dy_2^2\\[1ex]
+ dz_1^2 +dz_2^2 + R_t^2 \Big( d\tilde{x}_2^2 + \cdots + d \tilde{x}_5^2\Big)  \, , 
\end{multline}
and as usual, due to the NS-NS flux, the manifest~$SO(6)\times R^4$
symmetry is broken to~$SO(3)_y\times SO(3)_z \times R^4_{\tilde{x}}$.

\subsubsection{Penrose limit of the $D3$ brane}
\label{ses2}
We are now ready to discuss the Penrose limit of the $D3$
brane~(\ref{simple-cartezian}). Here we will use the following
parametrisation of the coordinates $(x_6,x_7,x_8,x_9)$ in~(\ref{Cartesian}):
\begin{equation}
x_7 = u \cos \psi \, , \, \, \,  x_8 = u \sin \psi \sin \xi \, , \, \, \,   x_6 = u \sin \psi \sin \xi \cos \eta 
\, ,  \, \,\,   x_9 = u \sin \psi  \sin \xi \sin \eta \, .
\end{equation}
Note that the parametrisation of the Poincar\'e coordinates which we
use here is different from the one used in equation~(\ref{cp}).
Other parameterisations correspond to different relative orientations
of the Penrose geode\-sic with respect to the brane worldvolume. In
principle these other choices could lead to different Penrose limits,
but it turns out that all these limits share the qualitative
characteristics of the one which we discuss.  Notice also that
if~$\alpha \neq 0$ the geodesic described by~(\ref{ged0})
and~(\ref{ged1}) cannot belong to the $D3$ brane for any value of the
parameters~$E,l,p_2$ and $p_3$, while if~$\alpha=0$ then
for~$p_2=p_3=0$ and arbitrary~$E,l$ the geodesic belongs to the brane
worldvolume.

To obtain the Penrose limit of the $D3$ brane we rewrite
equations~(\ref{simple-cartezian}), describing the embedding of $D3$ brane,
in the adapted coordinate system~(\ref{adapt}),
\begin{equation}
\label{brpl}
\begin{aligned}
&\tilde{x}_1 = \tilde{x}_4 = \tilde{x}_5 =  \tilde{\eta} = 0 \,, \\[1ex]
&{p_2 \over R_t^2}\, \tilde{u} + \tilde{x}_2 = {E \over L} \, \tan \, \alpha \, \sin\left({L \over R^2} \tilde{u}\right)\cos\left({l \over R^2} \tilde{u} + \tilde{\phi} - {p_2 \over l}\tilde{x}_2 - {p_3\over l} \tilde{x}_3\right)\,, \\[1ex]
&{p_3 \over R_t^2}\, \tilde{u} + \tilde{x}_3 = {E \over L} \tan\, \alpha\, \sin\left({L \over R^2} \tilde{u}\right)\, \sin\left({l \over R^2} \tilde{u} + \tilde{\phi} - {p_2 \over l} \tilde{x}_2 - {p_3 \over R^2} \tilde{x}_3\right) \sin\left({\pi \over 2}  + \tilde{\xi}\right)  \, . 
\end{aligned}
\end{equation}
We see that when~$p_2 \neq 0$ and~$p_3\neq 0$ an application of the
scaling~(\ref{scaling}) to the previous equations leads to equations
which can be satisfied only by~$\tilde{u}= 0$. Since~$\tilde{u}$ is a
time-like coordinate, this means that in the Penrose limit the~$D3$
brane becomes instantonic.

To obtain a more physical result, consider the~$p_2=p_3=0$ geodesic and
scale the angle~$\alpha$ such that
\begin{equation}
\tan \alpha \rightarrow \lambda \, \tan \alpha \, . 
\end{equation}
In other words: as we start zooming out the region near the geodesic
(which does not belong to the brane) we simultaneously bring the brane
closer and closer to the geodesic by sending the angle to zero.  In this
case equations~(\ref{brpl}) reduce in the limit to
\begin{equation}
\label{brwa}
\begin{aligned}
&y_2 = z_2 = \tilde{x}_4 = \tilde{x}_5 = 0 \, , \\[1ex]
&\tilde{x}_2 = \tan \, \alpha \, {E \over l } \sin\left({ 2 l \over R^2}
x^+\right)\cos\left({2 l \over R^2} x^+\right) \,, \\[1ex]
&\tilde{x}_3 = \tan\, \alpha\, {E \over l} \sin\left({2 l \over R^2}
x^+\right)\, \sin\left({2 l \over R^2} x^+\right)  \, .
\end{aligned}
\end{equation}
We see that the Penrose limit of the diagonal D3-brane produces a
brane with orientation~$(+,-,1,1,0)$, i.e.~a~$D3$ brane that extends
in the directions $x^+,x^-,y_1,z_1$. In addition, the brane carries a
null wave-like excitation in the directions~$\tilde{x}_2(x^+)$
and~$\tilde{x}_3(x^+)$.\footnote{We should emphasise that the wave
which is obtain here is a genuine wave, i.e.~it is not a mere
coordinate artifact, like the one which appeared
in~\cite{Bain:2002nq}.} These kind of wave excitations have been
studied recently in~\cite{Bachas:2002jg}. Note that the DBI action
allows for an arbitrary profile of transverse null-like
excitations.\footnote{It was shown in~\cite{Bachas:2002jg} that the
DBI waves are supersymmetric, \emph{exact} solutions of the DBI action
to all orders in~$\alpha'$.}  In our case, however, the precise form
of these worldvolume waves carries information about the position of
the brane with respect to the geodesic before the limit has been
taken. Note also that if we \emph{first} take the~$\alpha \rightarrow
0$ limit (and adjust the geodesic in such a way that it lies on the
worldvolume of the brane) and then take the Penrose limit, we would
end up with the~$(+,-,1,1,0)$ brane with no wave on its worldvolume.

\subsection{Penrose limit of the diagonal $D4$ brane: an oblique brane} 
\label{oblbr}

In this section we will study the Penrose limit of the diagonal~$D4$
brane~(\ref{dthb}).  We restrict to this simpler case, rather than
discussing the generic~$D4$ brane~(\ref{holbr}), since we are
interested in taking a Penrose limit along a geodesic that winds
between the two submanifolds and \emph{unlike} in the previous
section, \emph{belongs} to the brane.  The analysis of the geodesics
(which do or do not wind between submanifolds) but do not belong to
the brane worldvolume leads to Penrose limits similar to the one
which was already discussed.

So let us consider a geodesic which winds diagonally between the two
big circles of two three-spheres in the target space~(\ref{metr}) and
sits at the origin of the~$\text{AdS}_3$ space.  The product of two big
circles is just an (orthogonal) torus, and we focus on a geodesic
which diagonally winds in this torus.  If we use the
coordinates~(\ref{standard-s3}) for the three spheres, the geodesic
under the angle~$\omega$ (with respect to one of the big circles) can
be written as
\begin{equation}
\label{geod2}
\begin{aligned}
&\eta \equiv a \cos \omega  \eta_1 +  b \sin \omega \eta_2 = \tau  \, , \quad 
{\tilde  \eta} \equiv - a \sin \omega \, \eta_1 + b \cos \omega\, \eta_2 = c = \text{const}.\,,\\ 
&\rho = 0  \,,\, \,  \lambda = g = \text{const}. \quad \psi_i = \xi_i = {\pi \over 2} \quad (i=1,2)\, \quad  (a = {\sqrt{N_{F}^{(2)}} \over R} \, , \quad b = {\sqrt{N_F^{(1)}} \over R}) \, . 
\end{aligned}
\end{equation}
Going through the standard procedure \cite{Berenstein:2002jq} of introducing the light cone
coordinates $x^{\pm}$ and a set of new coordinates $(x_1,x_2,y_1,y_2,z_1,z_2, \phi, \tilde{\lambda})$ as
\begin{equation}
\label{scale}
\begin{aligned}
\tau &= x^+ + {1\over R^2} x^- \,,  & 
\eta &= x^+ - {1\over R^2} x^- \,,&
{\tilde \eta} &=  c + {1\over R} \phi \,, &
\lambda &= g + \tilde{\lambda} \,, \quad \rho = {\tilde{r}\over R} \,,\\[1ex]
\psi_1 &= {\pi \over 2}- {y_1 \over \sqrt{N_F^{(1)}}}  \,, &
\xi_1 &= {\pi \over 2}- { y_2 \over \sqrt{N_F^{(1)}}}  \,, &
\psi_2 &= {\pi \over 2}- { z_1 \over \sqrt{N_F^{(2)}}}  \,, &
\xi_2 &= {\pi \over 2}- {z_2 \over \sqrt{N_F^{(2)}}}  \,,
\end{aligned}
\end{equation}
with $\tilde{r}^2 = x_1^2 + x_2^2$, and scaling all parameters to
infinity in such a way that the parameters~$a$ and~$b$ are kept fixed, we
obtain the following pp-wave metric and three-form flux
\begin{equation}
\label{NW}
\begin{aligned}
ds^2 &\begin{aligned}[t]
  = - 4 dx^+ dx^-  - \Big(x_1^2 + x_2^2 + {\cos^2 \omega \over a^2}  
          (y_1^2 + y_2^2) + &{\sin^2 \omega \over b^2 }  (z_1^2 + z_2^2)\Big) (dx^+)^2 \\[1ex]
   &+ d {\vec x}^2 +  d \vec{y}^2  +  d \vec{z}^2 + d \phi^2 + d
  \lambda^2 \,, 
  \end{aligned}\\[2ex]
H &= 2 \, \tilde{r} dx^{+}\wedge d \tilde{r} \wedge d \sigma + 2 {\sin \omega \over b}\,  dx^{+} \wedge dy_1 \wedge dy_2 + 2 {\cos \omega \over a}  dx^{+} \wedge dz_1 \wedge dz_2 \, .
\end{aligned}
\end{equation}
This is just the standard Nappi-Witten background with different
values of NS flux in different two planes, and an isometry group (in
the space-like directions) given by the product \mbox{$G=SO(2)_{x}\times
SO(2)_{y}\times SO(2)_{z} \times R^2_{\lambda,\phi}$}.  The previous
discussion holds for arbitrary~$N_F^{(1)}$ and~$N_{F}^{(2)}$ but in
order to study~(\ref{dthb}) we furthermore restrict to the
case~$N_{F}^{(1)} = N_{F}^{(2)} \equiv N_{F}$.

Before we apply the Penrose limit to (\ref{dthb}), note that when
the parameters~$g,c$ and~$\omega$ take the values
\begin{equation}
g =R \ln(\tan \alpha)\,,  \quad c = 0 \, , \quad \omega = {\pi \over 4} \, ,
\end{equation}
the geodesic~(\ref{geod2}) satisfies the embedding equations of the
brane and hence belongs to its worldvolume.  Applying the
scaling~(\ref{scale}) to the equation~(\ref{dthb}), with the values of
the parameters given above, one obtains the embedding equations for
the $D4$ brane in the pp-wave~(\ref{NW}),
\begin{equation}
\label{diag-D4}
x_2 = y_1 = z_1 = 0  \, , \quad \phi = 0 \, , \quad {1 \over N_F} (z_2^2  -  y_2^2)  = {\lambda \over R}  \, .
\end{equation}
Since $\lambda$ does not scale, while $N_F$ and $R$ scale to infinity
in such a way that~$a=\text{const}.$, the third equation reduces in
the limit to~$\lambda =0$. In order to obtain a more interesting
Penrose limit we will scale~$\lambda$ to zero faster than~$1/R$. In
this case the third equation reduces to
\begin{equation}
\label{yz}
z_2 = \pm y_2 \, .
\end{equation}
We see that the resulting brane~(\ref{diag-D4}),~(\ref{yz}) has orientation
$(+,-,1,1/2,1/2,1)$. Here we have used the isometry group~$G$ to group
the brane worldvolume directions; the notation~$(1/2,1/2)$ means that
the brane is diagonally embedded between~$y_2$ and~$z_2$. Hence we 
have obtained a brane that is diagonally (i.e.~in an oblique way)
embedded between the~$SO(2)_y$ and~$SO(2)_z$ subspaces. This type of brane
was first observed in~\cite{Hikida:2002qk}, but their AdS origin was
not known so far.

Note however, that the scaling of~$\lambda$ which we have used leads
to an effective compactification in the limit (from ten to nine
dimensions). Although there are no conceptual reasons why one could
not take this kind of Penrose limit, this is not how the Penrose limit
is usually taken.  By considering a more complicated geodesic, one can
produce oblique branes from~(\ref{dthb}) without having to compactify
space in the limit. However, this computation is more tedious and
without an additional physical outcome, so we present it in the
appendix.

\subsection{Penrose limit of the brane with flux: a brane with null flux}
\label{pflux}

Let us now consider the Penrose limit of the~$\text{AdS}_4\times
S^4$,~$D7$ brane with diagonal worldvolume flux. We have seen in the
previous section that in order to obtain a nontrivial Penrose limit of
the brane under an angle, we were forced to bring the brane closer and
closer to the geodesic while taking the limit. Inspired by the
``equivalence'' of geometry and worldvolume fluxes, as in
section~\ref{seven}, we expect that a similar story will repeat itself
in the case of the brane with flux~(\ref{fflux}). There is however a
slight difference. In the purely geometrical setup, it is clear when
the geodesic lies along the brane, and hence when the ``standard''
Penrose limit (without any additional scaling of a parameter of the
solution) can be taken. In the case of a non-vanishing flux the
``standard'' geodesics should, as we will see, be taken in a direction
\emph{orthogonal} to the flux.

This feature can roughly be understood already in the context of~$3+1$
dimensional electromagnetism in flat space. Imagine that a magnetic
flux $F_{12}\,dx^1 \wedge dx^2$ is turned on. Boosting with the
velocity of light in the direction~$x^3$ results in zero electric and
magnetic fields (since we ``contract'' the length of the magnetic
field pointing in the third direction).  Boosting in the directions
one or two, on the other hand, leads to infinite electric and magnetic
fields.\footnote{Recall that for a general Lorentz transformation from
the system~$K$ to the system~$K'$ moving with velocity $\vec{v}$
relative to $K$, the transformation of the electric and magnetic
fields is given by
\begin{equation}
\begin{aligned}
\vec{E}' &= \gamma (\vec{E} + \vec{\beta} \times \vec{B}) - {\gamma^2 \over \gamma + 1} \vec{\beta}\cdot (\vec{\beta} \cdot \vec{E}) \,,\\
\vec{B}' &= \gamma (\vec{B} - \vec{\beta} \times \vec{E}) - {\gamma^2 \over \gamma + 1} \vec{\beta}\cdot (\vec{\beta} \cdot \vec{B}) \, .
\end{aligned}
\end{equation}} 
Taking the Penrose limit in curved space is a more complicated
procedure since it involves both boosting and additional rescaling,
so let us now check how this simple flat-space analysis
generalises to the full curved space.

Let us consider a geodesic that winds along the big circle
parametrised by $\eta$ and located at $\theta = \psi = \xi = \pi/2$ in
$S^5$ in~(\ref{S5}), and is located at the origin $\rho=0$ of $\text{AdS}_5$
in global coordinates. In Poincar\'e coordinates this geodesic is
lying in the~$4-5$ plane.  As usual, we now introduce new coordinates
and perform a rescaling as in~\cite{Berenstein:2002jq},
\begin{eqnarray}
\tau &=& x^+ + {x^-\over R^2} \, \quad \eta = x^+ - {x^-\over R^2} \, , 
\quad \rho = {z \over R} \, , \quad (z^2 = z_1^2 + \cdots + z_4^2) \,  , \nonumber \\
\theta &=& {\pi \over 2} + {\tilde{y} \over R} \, , \quad \psi = {\pi \over 2} + {y_3 \over R} \, , \quad \xi = {\pi \over 2} + {y_4 \over R} \, , \quad \varphi \rightarrow \varphi  \, \quad (\tilde{y}^2 = y_1^2 + y_2^2) \, .
\end{eqnarray}
In this limit the geometrical part of the~$D7$ brane reduces to a
brane with the embedding~$(+,-,3,3)$.

To see the effect of the Penrose limit on the flux, let us first
rotate the flux~(\ref{fflux}) in the $4-7$ and $5-8$ planes, so that
it is placed under arbitrary angles~$\nu_1$ and~$\nu_2$ with respect
to the above geodesic. For~$\nu_1 = \nu_2 = 0$ the flux is in the
plane $7-8$, i.e.~it is orthogonal to the geodesic. We will denote the
coordinates in section~\ref{seven} with primes and the coordinates
used in this section without primes. We have
\begin{equation}
\begin{aligned}
x_7' &= \cos \nu_1 \,x_7 + \sin \nu_1 \,x_4 \, , \quad  x_4' = - \sin \nu_1 \,x_7 + \cos \nu_1 \,x_4 \, , \\
x_8' &= \cos \nu_2 \,x_8 + \sin \nu_2 \,x_5 \, , \quad  x_5' = - \sin \nu_2 \,x_8 + \cos \nu_2 \,x_5 \, , \quad x_i' = x_i \, \, (\text{all other}) \, , 
\end{aligned}
\end{equation}
and the flux~(\ref{fflux}) becomes
\begin{equation}
\label{flux-angle}
\begin{aligned}
F &= - \cot \alpha\, dx_1' \wedge dx_2' + \tan \alpha\, dx_7' \wedge dx_8' \\[1ex]
  &= - \cot \alpha\,  dx_1 \wedge dx_2 + \tan \alpha \cos \nu_1 \cos \nu_2\, dx_7 \wedge dx_8 \\
&\quad + \tan \alpha \Big(\sin \nu_1 \cos \nu_2\, dx_4 \wedge dx_8  + \cos \nu_1 \sin \nu_2\, dx_7 \wedge dx_5 + \sin \nu_1 \sin \nu_2\, dx_4 \wedge dx_5\Big) \, .
\end{aligned}
\end{equation}
It is easy to see that the leading terms in the~$1/R$ expansion of the
flux are given by the following expression
\begin{equation}
\begin{aligned}
F = & -\cot\alpha \,  d\left({z_1\over \cos x^+}\right) \wedge d\left({z_2 \over \cos
  x^+}\right) \\[1ex]
&+ \tan \alpha  \cos \nu_1 \cos \nu_2 \Big( \tilde{y} \sin x^+ \cos x^+\, dx^+ \wedge d\varphi 
 - \cos^2 x^+\, d\tilde{y} \wedge d\varphi\Big)   \\[1ex]
&+ R\,\tan \alpha \tilde{y} \Big(\sin \nu_1 \cos \nu_2  \cos 2 x^+ \,+ \ \cos \nu_1 \sin \nu_2  \sin 2 x^+ \Big) \cos x^+ \, \cos \varphi\, d \varphi \wedge dx^+ \\ 
&- R\,\tan \alpha \sin \nu_1 \sin \nu_2 \cos x^+\, dx^+ \wedge d z_4 \,
 + {\cal O}\left({1\over R}\right),
\end{aligned}
\end{equation}
where $z_4 \equiv z \Omega_4$ and the angular parameter~$\Omega_4 $ is
defined in~(\ref{phic}).  We see that if the Penrose geodesic is taken
to be orthogonal to the flux~(i.e.~$\nu_1 = \nu_2 = 0$) then there is
no need to perform any additional scaling of the above flux. If
however~$\nu_1 \neq 0$ and~$\nu_2 \neq 0$, then for the Penrose limit
to be well-defined we have to simultaneously send~$\nu_1 \rightarrow
0$ and~$\nu_2 \rightarrow 0$ while taking the limit. As in the purely
geometrical case, taking the Penrose limit and rescaling the
parameters are operations that do not commute. The flux obtained by
taking the Penrose limit along the orthogonal geodesic is
\emph{different} from the one obtained by looking at a geodesic under
an angle with the angle rescaled to zero while taking the limit.

The case of an orthogonal flux has already appeared in the
literature~\cite{Skenderis:2002vf} in the case of the Penrose limit of
the~$\text{AdS}_4 \times S^2$ brane which wraps a maximal~$S^2$ and carries
magnetic flux on the~$S^2$. In this case the Penrose limit was taken
along the~$S^1$ in~$S^2$ i.e.~parallel to the flux, with the expected
consequence that the flux had to be rescaled in order to obtain a
finite Penrose limit.

Finally let us conclude this section with the observation that the
brane with flux which was obtained by considering the previous Penrose
limit is of the form discussed recently in~\cite{Durin:2003gj}. In
flat space a D-brane with a null flux is T-dual to a D-brane with a
null pulse, like the one which was obtained in the previous section.
In the pp-wave this duality is less manifest, since T-duality changes
the background as well. Throughout this paper we have tried to trace
how these dual configurations are realised in flat space, AdS and
finally pp-wave spacetimes.

\section{Discussion}

In this paper we have presented various types of D-branes which mix
target space manifolds in a geometrical way or through diagonal
worldvolume fluxes. Our consideration was classical and based on a
probe brane analysis. It would be interesting to extend our analysis
to the supergravity regime, taking into account the back-reaction of
the D-branes as in~\cite{Bain:2002nq}. It would also be desirable to
obtain a description of these branes (at least in the Penrose limit)
using the covariant approach of~\cite{Bain:2002tq}.

Our initial motivation for studying diagonal D-branes in product
spaces was to understand the construction of boundary states for these
branes. This problem, however, turned out to be hard, essentially due
to the fact that some of the target space symmetries are completely
broken by the brane; there is no diagonal current preserved by the
boundary conditions. This of course does not mean that the brane is
inconsistent. The consistency of the theory only requires the diagonal
part of the Virasoro algebra to be preserved. Moreover, due to the way
these branes were constructed, we are guaranteed that they are
consistent. However, it is not clear to us at the moment how one
should implement the properties of these branes at the level of
boundary states.  One possible way in which one might be able to
improve the understanding of some of the features of diagonal branes
is by a classical analysis of their spectrum (as
in~\cite{Bachas:2000ik}). An alternative way, mentioned in the
introduction, is to study these branes using the dCFT/AdS
correspondence~\cite{sarkissian}. Finally, a construction of the
boundary states in the Penrose limit might also help to understand
these branes in AdS geometries.

\section*{Acknowledgements}
We would like to thank Matthias Blau, Jerome Gauntlett, George Papadopoulos, Kasper
Peeters and Volker Schomerus for useful discussions.

\appendix
\section{Technical details}
\subsection{Technical details about the Penrose limit of $\text{AdS}_3 \times S^3 \times T^4$}

In this section we give some technical details about the calculation
in section~\ref{ses1}.  One possible choice for the adapted coordinate
system which fulfills the requirements listed around~(\ref{newc}) is
given by
\begin{equation}
\label{adapt}
\begin{aligned}
u &= {E \over L} \, \sin\left({L \over R^2} \tilde{u}\right) \, , & &\psi = {l \over R^2} \, \tilde{u} + \tilde{\phi} - {p_2 \over l} \, \tilde{x}_2 - {p_3 \over l}\,  \tilde{x}_3 \, , \\
t &= - {L \over E} \, \cot \left({L \over R^2} \tilde{u}\right) - {R^2
  \over E} \, \tilde{v}  +  {l \over E} \, \tilde{\phi} \, , & & x_2 = {p_2 \over R_t^2} \, \tilde{u} + \tilde{x}_2 \, , \quad x_3 = {p_3 \over R_t^2} \, \tilde{u} + \tilde{x}_3 \,,\\
\xi &= {\pi \over 2} + \tilde{\xi} \,, \quad  \eta = \tilde{\eta} \, ,
\quad x_i = \tilde{x}_i \, , & & \text{with}\quad i=1,4,5 \, .
\end{aligned}
\end{equation}
The metric~(\ref{Poincare}) in adapted coordinates becomes
\begin{equation}
\begin{aligned}
\label{rose1}
ds^2   =& 2 R^2 \, d\tilde{u} d\tilde{v} - {R^6 \over L^2}\,
\sin^2\left({L \over R^2} \tilde{u}\right) d\tilde{v}^2 
+ \left(-{l^2 \over L^2} R^2\,
\sin^2\left({L \over R^2} \tilde{u}\right) +  R^2\right) d \tilde{\phi}^2 \\[1ex]
&+ 2 R^4 {l \over L^2} \sin^2\left({L \over R^2} \tilde{u}\right) d \tilde{v} d\tilde{\phi} 
- 2 {p_2 \over l} R^2 d \tilde{x}_2 d \tilde{\phi} - 2 {p_3 \over l} R^2 d \tilde{x}_3 d \tilde{\phi}  \\[1ex]
&+ \Big(R_t^2 + {p_2^2  \over l^2} R^2\Big) d\tilde{x}_2^2 
+ \Big(R_t^2 + {p_3^2
  \over l^2} R^2\Big) d\tilde{x}_3^2 
+ {2p_2p_3R^2\over l^2}d\tilde{x}_2d\tilde{x}_3 \\[1ex]
& + R_t^2 \Big(d\tilde{x}_4^2 + d\tilde{x}_5^2\Big) + R^2 \sin^2(\psi)\,
 ds^2(S^2) 
+ {R^2 E^2 \over L^2} \sin^2\left({L \over R^2} \tilde{u}\right) d\tilde{x}_1^2 \, . 
\end{aligned}
\end{equation}
Next we apply the scaling~(\ref{scaling}) to the metric~(\ref{rose1})
and set~$p_2 = p_3 = 0$. In order to rewrite the metric in Brinkman
coordinates, one then uses the following coordinate
transformations~\cite{Blau:2002mw}:
\begin{equation}
\label{change}
\begin{aligned}
\tilde{u} &= 2 x^+ \, , \quad \tilde{v} = {1\over 2 R^2}\left(x^- + {1\over 2} \sum_{i =
1}^{4} a_i^{-1} \partial_{x^+}(a_i) (x_i)^2  \right) \, ,  \quad (x_i \equiv y_1,y_2,z_1,z_2) \, , \\[.5ex]
y_1 &= R \cos \left({2 L\over R^2} x^{+}\right) \tilde{\varphi} \equiv a_1(x^+)
\tilde{\varphi} \, , \quad \quad\smash{\begin{aligned}[t]
y_2 &= {R E \over L} \sin \left({2 L\over R^2} x^{+}\right) \tilde{x}_1 \equiv a_2(x^+)
\tilde{x}_1 \, , \\
z_2 &= R \sin \left({2 L\over R^2} x^{+}\right) \tilde{\eta} \equiv a_4(x^+) \tilde{\eta} \,\,,
\end{aligned}}\\
z_1 &= R \sin \left({2 L\over R^2} x^{+}\right) \tilde{\xi} \equiv a_3(x^+)
\tilde{\xi} \, ,
\end{aligned}
\end{equation}
while the remainder of the coordinates remain unchanged. The resulting
metric is given in equation~(\ref{limi}).

\subsection{Technical details about the Penrose limit of the diagonal D4-brane}

Here we present a Penrose limit of the
diagonal~$D4$ brane~(\ref{dthb}) which is different from the one given
in section~\ref{oblbr}. The resulting brane is again an oblique brane,
but the target space is ten-dimensional, unlike the one
in~(\ref{diag-D4}) and~(\ref{yz}).  Let us consider the following
geodesic in Poincar\'e coordinates,
\begin{eqnarray}
\label{geod3}
u &=& {E \over L} \sin \left( {L \over R^2} \tau \right) \, , \quad t = - {L \over E} \cot \left(  {L \over R^2} \tau  \right)  \, , \nonumber \\
\xi_a &\equiv& \xi_1 +   \xi_2 = \tau  \, , \quad 
\xi_b \equiv -   \xi_1 +  \xi_2 = c_\xi .    \nonumber \\ 
\psi_1 &=& \psi_2 = {\pi \over 2}  \,,\, \,  \eta_1 = \eta_2 = c_\eta \, , \quad \lambda = g \, , \, \,  x_1 = c_{x} \, ,\, \,  a =b= {\sqrt{N_{F}} \over R}=\sqrt{2} \, . 
\end{eqnarray}
It is easy to see that for~$c_{x} = c_\eta = c_\xi = g = 0$, this
geodesic belongs to the worldvolume of the brane. It winds along a
diagonal curve ($\xi_1= \xi_2$) which is ``orthogonal'' to the other
diagonal direction ($\eta_1=\eta_2$) that the brane wraps as well. In
what follows we will make a restriction to this kind of geodesic.

Next we go to coordinate system adapted to this geodesic, following
a procedure similar to the one in section~\ref{plimt}.
\begin{equation}
(u,t, x_1 ,\psi_1, \psi_2, \xi_a, \xi_b, \eta_1, \eta_2) \rightarrow ( \tilde{u},\tilde{v}, \tilde{x}_1, \tilde{\psi}_1, \tilde{\psi}_2, \xi_1, \tilde{\xi}_a, \tilde{\eta}_a, \tilde{\eta}_b ) \, ,
\end{equation}
where the new coordinates are given by
\begin{eqnarray}
u &=& {E \over L} \sin \left( {L \over R^2} \tilde{u} \right) \, , \quad t = - {L\over E} \cot \left(  {L \over R^2} \tilde{u}  \right) - {R^2 \over E} \tilde{v} + {L\over E} \tilde{\xi}_a  \, , \nonumber \\
\xi_a &=& {L \over R^2} \tilde{u} + \tilde{\xi}_a \, ,\quad \xi_b = \tilde{\xi}_b \, \nonumber \\
\psi_1 &=& {\pi \over 2} + \tilde{\psi}_1 \, , \quad \psi_2 = {\pi \over 2} + \tilde{\psi}_2 \, ,   \nonumber \\
\tilde{\lambda} &=& \lambda,  \quad  \tilde{x}_1 = x_1 \,  .
\end{eqnarray}
The $\text{AdS}_3\times S^3 \times S^3 \times R$ metric written in  these
coordinates is given by
\begin{eqnarray}
ds^2 &=& \left( {L \over R} \right)^2  \left(\cos^2 \tilde{\psi}_1  + \cos^2 \tilde{\psi}_2  - 2 \right) {d\tilde{u}^2\over 2}  + 2R^2 d\tilde{u} d \tilde{v}  - {R^6 \over L^2} \sin^2\left({L \over R^2} \tilde{u}\right) d \tilde{v}^2 \nonumber\\
&+& L (-2  + \cos^2 \tilde{\psi}_1 + \cos^2 \tilde{\psi}_2) d\tilde{u}d\tilde{\xi}_a  
 +  L(\cos^2 \tilde{\psi}_2 - \cos^2 \tilde{\psi}_1) d\tilde{u} d \tilde{\xi}_b \nonumber\\
&+& R^2 \Big(-\sin^2 ( {L \over R^2} \tilde{u})
 +  {\cos^2\tilde{\psi}_1+\cos^2\tilde{\psi}_2\over 2}\Big)d \tilde{\xi}_a^2  + 
{R^2\over 2}(\cos^2\tilde{\psi}_1+\cos^2\tilde{\psi}_2) d \tilde{\xi}_b^2 \nonumber \\
&+& R^2 \Big(\cos^2 \tilde{\psi}_2 - \cos^2 \tilde{\psi}_1\Big) d \xi_a d \xi_b \, 
+ 2R^2 \Big(d \tilde{\psi}_1^2 +  d \tilde{\psi}_2^2\Big) \\
&+& 2R^2 \cos^2 \tilde{\psi}_1\sin^2 {1\over 2}\left( {L \over R^2} \tilde{u}+
\tilde{\xi}_a-\tilde{\xi}_b\right)d\tilde{\eta}_1^2\nonumber\\
& +& 2R^2 \cos^2 \tilde{\psi}_2\sin^2 {1\over 2}\left( {L \over R^2} \tilde{u}+
\tilde{\xi}_a+\tilde{\xi}_b\right)d \tilde{\eta}_2^2 \nonumber\\
& +& R^2 {E^2 \over L^2} \sin^2 \Big({L \over R^2} \tilde{u}\Big) d \tilde{x}_1^2
+{R^4\over L}\sin^2\Big({L\tilde{u}\over R^2}\Big)d\tilde{v}d\tilde{\xi}_a +  d\tilde{\lambda}^2\, .\nonumber
\end{eqnarray}
We then apply the following rescaling of the metric, using~$\Lambda
\rightarrow 0$,
\begin{eqnarray}
\label{scaling2}
&&R \rightarrow \Lambda^{-1} R\, , \quad   \tilde{u} \rightarrow \Lambda^{-2} \tilde{u} \, , \quad \tilde{v} \rightarrow \Lambda^4  \tilde{v} \, , \quad  \tilde{\lambda} \rightarrow \tilde{\lambda}  \nonumber \\
&&\tilde{x}_i \rightarrow \Lambda \,\tilde{x}_i \, , \quad (\tilde{x}_i = \tilde{\psi}_{1},\tilde{\psi}_2, \tilde{\eta}_{1}, \tilde{\eta}_2, \tilde{\xi}_{a},\tilde{\xi}_b , \tilde{x}_1 \, ) \, .
\end{eqnarray}
Note that the coordinate~$\tilde{\lambda}$ does not get scaled in the
limit.  Under this scaling the metric becomes
\begin{equation}
\begin{aligned}
ds^2 = {L^2 \over R^4} & (- \tilde{\psi}_1^2 - \tilde{\psi}_2^2) {d \tilde{u}^2\over 2}  + 2R^2 d \tilde{u} d \tilde{v} + R^2 \cos^2\left( {L \over R^2} \tilde{u} \right) d \tilde{\xi}_a^2 + R^2 d \tilde{\xi}_b^2  \\
&+  2R^2 (d\tilde{\psi}_1^2 +  d\tilde{\psi}_2^2)  + 2R^2 \sin^2
\left({L \over 2R^2} \tilde{u}\right) (d\tilde{\eta}_1^2 +
d\tilde{\eta}_2^2) \\
& + R^2 {E^2 \over L^2} \sin^2 \left({L \over R^2}
\tilde{u}\right) d \tilde{x}_1^2 + d\tilde{\lambda}^2\, \, .
\end{aligned}
\end{equation}
To put this metric into Brinkman form we have to make a change of
coordinates as in the previous section,
\begin{equation}
\begin{aligned}
\tilde{v} &= {1\over 2 R^2} \left( x^- + {1\over 2} \sum_{i =
1}^{4} a_i^{-1} \partial_{x^+}(a_i) (y^i)^2 \right)  \, ,  &
\hat{\eta}_{1,2} &= R \sin \left({L \over 2R^2} \tilde{u}\right) \tilde{\eta}_{1,2} \, \equiv a_i (u) y_i \, ,  \\
 \tilde{u} &= 2 x^+ \, , \quad \hat{\xi}_a = R \cos \left({L \over
	R^2} \tilde{u}\right) \tilde{\xi}_a \, \equiv a_3 (u) y_3 \,, & \hat{x}_1 &= R {E \over L}  \sin \left({L \over R^2} \tilde{u}\right) \tilde{x}_1 \equiv a_4 (u) y_4  \, .
\end{aligned}
\end{equation}
and all other coordinates are kept unchanged.  So in Brinkman coordinates the
metric becomes
\begin{equation}
ds^2 = 2 dx^+ dx^- - {L^2 \over R^4} \left(  2\tilde{\psi}_1^2 + 2\tilde{\psi}_2^2  + 4\hat{\xi}_a^2  + \hat{\eta}_1^2  + \hat{\eta}_2^2  + 4\hat{x}_1^2 \right) (dx^+)^2 + dE_8^2\, .
\end{equation}
Under the above scaling and after the change of coordinates the
equations for the embedding of the brane become
\begin{eqnarray}
\tilde{\psi}_1 &=& \tilde{\psi}_2 = 0 \, \nonumber \\ 
\hat{\eta}_1 &=& \hat{\eta}_2  \, \nonumber \\
\lambda &=& - 2R \cot \left(   {L \over R^2} x^{+}\right)\hat{\xi}_b  \, .
\end{eqnarray}
Hence we see that the limiting brane is oblique, as advertised.

\subsection{Various coordinate systems and relations between them}

\underline{$\text{AdS}_{p+2}$}
\par\nopagebreak\vskip5pt
\noindent The~$p+2$-dimensional~AdS space can be viewed as a hyperboloid in
$p+3$-dimensional \emph{flat} space with a Lorentzian metric of
signature~$(-,-,+,\dots,+)$,
\begin{equation}
\label{hyper}
X_{0}^2 + X_{p+2}- X_1^2 \cdots -X_{p+1}^2 = R^2 \, .
\end{equation}
In this paper we use the following two parametrisations of this
hyperboloid:
\begin{eqnarray}
\label{parm}
&&X_0 = R \cosh \rho \cos \tau ={1\over 2} \left({1\over u}+ u(R^2+ \vec{x}^2 - t^2)\right) \, , \quad \vec{x} \in R^p \, , \nonumber \\
&&X_{p+2} = R \cosh \rho \sin \tau = R u t \, , \nonumber \\
&&X_i = R \sinh \rho \Omega_i = R u x_i \,  , \quad (i=1\dots p) \, , \nonumber \\ 
&&X_{p+1} = R \sinh \rho \Omega_{p+1} = {1\over 2 u} (1 - u^2 (R^2 - \vec{x}^2 + t^2))\, , \quad   \sum_{i=1}^{p+1}  \Omega_i^2 =R^2 \,.
\end{eqnarray}
The induced metric on the hyperboloid, written using the second
parametrisation, leads to the metric in Poincar\'e
coordinates~(\ref{Poincare}), while the first parametrisation leads to
the AdS metric in global coordinates
\begin{equation}
\label{AdS-global}
ds^2 = R^2\big(-\cosh^2\rho d\tau^2 + d \rho^2 + \sinh^2 \rho d\Omega_{S^p}^2\big) \, .
\end{equation}
For the $\text{AdS}_3$ space we have $p=1$ and we use $\Omega_1 = \cos \sigma$. 
\bigskip

\noindent \underline{$\text{AdS}_3$}
\par\nopagebreak\vskip5pt
\noindent The $\text{AdS}_3$ space is the (universal cover of the) group manifold of the
non-compact group $SL(2,\mathbb{R})$. A generic element in this group can be
written as
\begin{equation}
\label{SL2R}
g = {1\over R} \begin{pmatrix}
X_0 + X_1 \, & \, X_2 + X_3 \\
X_2 - X_3  \, &  \, X_0 - X_1 
\end{pmatrix}
\end{equation}
subject to condition that the determinant of~(\ref{SL2R}) is equal to
one, which is precisely the equation of the hyperboloid given in~(\ref{hyper}).
\bigskip

\noindent \underline{$S^3$} 
\par\nopagebreak\vskip5pt
\noindent In the main text we use three different ways to write the metric
on~$S^3$. Firstly, using the Euler parametrisation of the group element
we have
\begin{eqnarray}
\label{euler-s3}
g &=& e^{i\chi {\sigma_3 \over 2}}e^{i \tilde{\theta} {\sigma_1 \over 2}} e^{i \varphi {\sigma_3 \over 2}} \, \nonumber \\ 
ds^2 &=& {1\over 4} [ (d \chi + \cos \tilde{\theta} d \varphi)^2 + d \tilde{\theta}^2 + \sin^2 \tilde{\theta} d \varphi^2 ] \, .
\end{eqnarray}
Secondly, we can use coordinates which are analogue to the global
coordinate for~$\text{AdS}_3$,
\begin{eqnarray}
\label{global-s3}
X_1+iX_2 &=& \cos \theta e^{i\tilde{\phi}} \, , \quad X_3+iX_4 = \sin \theta e^{i\phi}\, \nonumber \\ 
ds^2 &=& d \theta^2 + \cos^2 \theta d \tilde{\phi}^2 + \sin^2 \theta d \phi^2 \, .  
\end{eqnarray}
The relation between the metrics~(\ref{euler-s3}) and~(\ref{global-s3}) is given by
\begin{equation}
\label{glob-par}
\chi = \tilde{\phi} + \phi  \, , \quad \varphi = \tilde{\phi} - \phi \, ,
 \quad \theta={\tilde{\theta} \over 2} 
\end{equation}
Here the ranges for~$\tilde{\phi}$, $\phi$ and $\theta$ are
$-\pi\leq\tilde{\phi},\phi\leq\pi$ and $0\leq\theta\leq{\pi\over 2}$
respectively.

Thirdly, the standard metric on~$S^3$ can be written as (using the
unit vector $\vec{n}$ on $S^2$)
\begin{eqnarray}
\label{standard-s3}
g &=& e^{2i  \psi {\vec{n} \cdot \vec{\sigma} \over 2}} \, , \quad
ds^2 = d \psi^2 + \sin^2 \psi ( d \xi^2 + \sin^2 \xi d \eta^2) \, \nonumber \\
X_1+iX_2&=& \cos \psi +i\sin \psi\cos \xi \, , \quad X_3+iX_4 = \sin \psi \sin \xi e^{i\eta} 
\end{eqnarray}
Here the range of coordinates is $ 0 \leq \psi, \xi \leq \pi $ and
$0 \leq\ \eta < 2\pi$.

\subsection{General equations of motion for the DBI action}

The general D$p$ brane equations of motion, derived
in~\cite{Skenderis:2002vf}, are given by
\begin{equation}
\label{general1}
\partial_{i} (\sqrt{- \det M} \theta^{ii_1}) = \epsilon^{i_1 \cdots
  i_{p+1}} \sum_{n\geq 0} {1\over n! \, (2!)^n\, (q-1)!}({\cal
  F})^n_{i_2 \cdots i_{2n+1}} \bar{F}_{i_{2n+2}\cdots i_{p+1}} \,,
\end{equation}
\begin{multline}
\label{general2}
\sum_{n\geq 0} {1\over n! \, (2!)^n\, q!} \epsilon^{i_1 \cdots i_{p+1}}({\cal F})^n_{i_1 \cdots i_{2n}} \bar{F}_{\mu i_{2n+1}\cdots i_{p+1}} \\
= e^{\Phi} \bigg{(}  \sqrt{-M}(G^{ij} \partial_i X^{\nu}\partial_j X^{\xi} g_{\mu\nu} \Phi_{,\xi} - \Phi_{,\mu})
- {\cal K}_{\mu} \bigg{)} 
\end{multline}
where
\begin{equation}
{\cal K}_{\mu} = - \partial_i(\sqrt{- \det M} G^{ij})\partial_j\, X^{\nu} g_{\mu\nu} -  \sqrt{ - \det M} M^{ij} \bigg{(} \partial_i\partial_j X^{\nu} g_{\mu\nu} + \tilde{\Gamma}_{\mu\nu\xi} \partial_{i} X^{\nu} \partial_j X^\xi \bigg{)}  \, ,   
\end{equation}
and $\bar{F}$ is the pull-back of the target space RR
fields~$C_{[q]}$\footnote{Only the $\mu$ index in~(\ref{general2}) is
not a pulled-back index.}
\begin{equation}
\bar{F}_{\mu_1 \cdots \mu_{q+1}} = (q+1)\partial_{[\mu_1}\,C_{\mu_2 \cdots \mu_{q+1}]} - {(q+1)! \over 3! (q-2)!}\, H_{[\mu_1\mu_2\mu_3}\, C_{\mu_4 \cdots \mu_{(q+1)]}} \, .
\end{equation}
The other quantities appearing in the above formulae are given by
\begin{equation}
\label{MT}
M_{ij} = \partial_i X^{\mu} \partial_j X^\nu g_{\mu \nu} + {\cal F}_{ij} \, , \quad  \theta^{ij}\equiv M^{[ij]} \, ,\quad G^{ij} \equiv M^{(ij)} \,,  \quad M^{ik} M_{kj} = \delta^{i}_{j} \,   .
\end{equation}
and $\tilde{\Gamma}$ is torsionful connection $\tilde{\Gamma} =
\Gamma - {1\over 2} H$. Also $\cal{F}$ is the gauge invariant two-form
\begin{equation}
{\cal F}_{ij} = F_{ij} - \partial_i X^{\mu}\partial_j X^{\nu} B_{\mu\nu} \, .
\end{equation}

\bibliographystyle{JHEP} 
\bibliography{diagonal}

\providecommand{\href}[2]{#2}\begingroup\raggedright\begin{thebibliography}{10}

\bibitem{Bilal:1998ck}
A.~Bilal and C.-S. Chu, {\it D3 brane(s) in {AdS$_5 \times S^5$} and
  {$N=4,2,1$} {SYM}},  {\em Nucl. Phys.} {\bf B547} (1999) 179--200,
[\href{http://xxx.lanl.gov/abs/hep-th/9810195}{{\tt hep-th/9810195}}].

\bibitem{Gutowski:1999iu}
J.~Gutowski and G.~Papadopoulos, {\it {AdS} calibrations},  {\em Phys. Lett.}
  {\bf B462} (1999) 81--88,
[\href{http://xxx.lanl.gov/abs/hep-th/9902034}{{\tt hep-th/9902034}}].

\bibitem{Gutowski:1999tu}
J.~Gutowski, G.~Papadopoulos, and P.~K. Townsend, {\it Supersymmetry and
  generalized calibrations},  {\em Phys. Rev.} {\bf D60} (1999) 106006,
[\href{http://xxx.lanl.gov/abs/hep-th/9905156}{{\tt hep-th/9905156}}].

\bibitem{sarkissian}
G.Sarkissian and M.~Zamaklar,
{\it in preparation}, .

\bibitem{Gibbons:1998hm}
G.~W. Gibbons and G.~Papadopoulos, {\it Calibrations and intersecting branes},
  {\em Commun. Math. Phys.} {\bf 202} (1999) 593--619,
[\href{http://xxx.lanl.gov/abs/hep-th/9803163}{{\tt hep-th/9803163}}].

\bibitem{Bachas:2000fr}
C.~Bachas and M.~Petropoulos, {\it {Anti-de-Sitter D-branes}},  {\em JHEP} {\bf
  02} (2001) 025,
[\href{http://xxx.lanl.gov/abs/hep-th/0012234}{{\tt hep-th/0012234}}].

\bibitem{Karch:2000ct}
A.~Karch and L.~Randall, {\it Locally localized gravity},  {\em JHEP} {\bf 05}
  (2001) 008,
[\href{http://xxx.lanl.gov/abs/hep-th/0011156}{{\tt hep-th/0011156}}].

\bibitem{Bachas:2002jg}
C.~Bachas, {\it Relativistic string in a pulse},  {\em Ann. Phys.} {\bf 305}
  (2003) 286--309,
[\href{http://xxx.lanl.gov/abs/hep-th/0212217}{{\tt hep-th/0212217}}].

\bibitem{Hikida:2002qk}
Y.~Hikida and S.~Yamaguchi, {\it D-branes in pp-waves and massive theories on
  worldsheet with boundary},  {\em JHEP} {\bf 01} (2003) 072,
[\href{http://xxx.lanl.gov/abs/hep-th/0210262}{{\tt hep-th/0210262}}].

\bibitem{Gaberdiel:2003sb}
M.~R. Gaberdiel, M.~B. Green, S.~Schafer-Nameki, and A.~Sinha, {\it Oblique and
  curved {D-branes} in {IIB} plane-wave string theory},  {\em JHEP} {\bf 10}
  (2003) 052,
[\href{http://xxx.lanl.gov/abs/hep-th/0306056}{{\tt hep-th/0306056}}].

\bibitem{Gauntlett:1997cv}
J.~P. Gauntlett, {\it Intersecting branes},
\href{http://xxx.lanl.gov/abs/hep-th/9705011}{{\tt hep-th/9705011}}.

\bibitem{HL}
R.Harvey and H.~Lawson, {\it Calibrated geometries},  {\em Acta Math.} {\bf
  148} (1982) 47.

\bibitem{H}
R.Harvey, {\it Spinors and calibrations},  {\em Acta Math.} {\bf 148} (1982)
  47.

\bibitem{Hanany:1997ie}
A.~Hanany and E.~Witten, {\it {Type IIB} superstrings, {BPS} monopoles, and
  three-dimensional gauge dynamics},  {\em Nucl. Phys.} {\bf B492} (1997)
  152--190,
[\href{http://xxx.lanl.gov/abs/hep-th/9611230}{{\tt hep-th/9611230}}].

\bibitem{Karch:2002sh}
A.~Karch and E.~Katz, {\it Adding flavor to {AdS/CFT}},  {\em JHEP} {\bf 06}
  (2002) 043,
[\href{http://xxx.lanl.gov/abs/hep-th/0205236}{{\tt hep-th/0205236}}].

\bibitem{Seiberg:1999vs}
N.~Seiberg and E.~Witten, {\it String theory and noncommutative geometry},
  {\em JHEP} {\bf 09} (1999) 032,
[\href{http://xxx.lanl.gov/abs/hep-th/9908142}{{\tt hep-th/9908142}}].

\bibitem{Cowdall:1998bu}
P.~M. Cowdall and P.~K. Townsend, {\it Gauged supergravity vacua from
  intersecting branes},  {\em Phys. Lett.} {\bf B429} (1998) 281--288,
[\href{http://xxx.lanl.gov/abs/hep-th/9801165}{{\tt hep-th/9801165}}].

\bibitem{Gauntlett:1998kc}
J.~P. Gauntlett, R.~C. Myers, and P.~K. Townsend, {\it Supersymmetry of
  rotating branes},  {\em Phys. Rev.} {\bf D59} (1999) 025001,
[\href{http://xxx.lanl.gov/abs/hep-th/9809065}{{\tt hep-th/9809065}}].

\bibitem{deBoer:1999rh}
J.~de~Boer, A.~Pasquinucci, and K.~Skenderis, {\it {AdS/CFT} dualities
  involving large 2d {$N=4$} superconformal symmetry},  {\em Adv. Theor. Math.
  Phys.} {\bf 3} (1999) 577--614,
[\href{http://xxx.lanl.gov/abs/hep-th/9904073}{{\tt hep-th/9904073}}].

\bibitem{Skenderis:2002vf}
K.~Skenderis and M.~Taylor, {\it Branes in {AdS} and pp-wave spacetimes},  {\em
  JHEP} {\bf 06} (2002) 025,
[\href{http://xxx.lanl.gov/abs/hep-th/0204054}{{\tt hep-th/0204054}}].

\bibitem{Marino:1999af}
M.~Marino, R.~Minasian, G.~W. Moore, and A.~Strominger, {\it Nonlinear
  instantons from supersymmetric p-branes},  {\em JHEP} {\bf 01} (2000) 005,
[\href{http://xxx.lanl.gov/abs/hep-th/9911206}{{\tt hep-th/9911206}}].

\bibitem{Blau:2002mw}
M.~Blau, J.~Figueroa-O'Farrill, and G.~Papadopoulos, {\it Penrose limits,
  supergravity and brane dynamics},  {\em Class. Quant. Grav.} {\bf 19} (2002)
  4753,
[\href{http://xxx.lanl.gov/abs/hep-th/0202111}{{\tt hep-th/0202111}}].

\bibitem{Bain:2002nq}
P.~Bain, P.~Meessen, and M.~Zamaklar, {\it Supergravity solutions for
  {D-branes} in {Hpp-wave} backgrounds},  {\em Class. Quant. Grav.} {\bf 20}
  (2003) 913--934,
[\href{http://xxx.lanl.gov/abs/hep-th/0205106}{{\tt hep-th/0205106}}].

\bibitem{Berenstein:2002jq}
D.~Berenstein, J.~M. Maldacena, and H.~Nastase, {\it Strings in flat space and
  pp waves from {$N=4$} super {Yang-Mills}},  {\em JHEP} {\bf 04} (2002) 013,
[\href{http://xxx.lanl.gov/abs/hep-th/0202021}{{\tt hep-th/0202021}}].

\bibitem{Durin:2003gj}
B.~Durin and B.~Pioline, {\it Open strings in relativistic ion traps},  {\em
  JHEP} {\bf 05} (2003) 035,
[\href{http://xxx.lanl.gov/abs/hep-th/0302159}{{\tt hep-th/0302159}}].

\bibitem{Bain:2002tq}
P.~Bain, K.~Peeters, and M.~Zamaklar, {\it D-branes in a plane wave from
  covariant open strings},  {\em Phys. Rev.} {\bf D67} (2003) 066001,
[\href{http://xxx.lanl.gov/abs/hep-th/0208038}{{\tt hep-th/0208038}}].

\bibitem{Bachas:2000ik}
C.~Bachas, M.~R. Douglas, and C.~Schweigert, {\it Flux stabilization of
  {D-branes}},  {\em JHEP} {\bf 05} (2000) 048,
[\href{http://xxx.lanl.gov/abs/hep-th/0003037}{{\tt hep-th/0003037}}].

\end{thebibliography}\endgroup

\end{document}